\documentclass[letterpaper]{article} 
\usepackage{aaai23}  
\usepackage{times}  
\usepackage{helvet}  
\usepackage{courier}  
\usepackage[hyphens]{url}  
\usepackage{graphicx} 
\usepackage{natbib}  
\usepackage{caption} 
\usepackage{algorithm}
\usepackage{algorithmic}
\usepackage{import}
\usepackage{subfig}
\usepackage{amsmath}
\DeclareMathOperator*{\argmax}{arg\,max}

\usepackage{amsfonts}
\usepackage{amssymb}
\usepackage{pifont}
\newcommand{\cmark}{\ding{51}}
\newcommand{\xmark}{\ding{55}}
\usepackage{array}
\newcolumntype{P}[1]{>{\centering\arraybackslash}p{#1}}
\newcolumntype{M}[1]{>{\centering\arraybackslash}m{#1}}
\usepackage{xcolor}
\usepackage{soul}
\usepackage{newfloat}
\usepackage{listings}
\DeclareCaptionStyle{ruled}{labelfont=normalfont,labelsep=colon,strut=off} 
\floatstyle{ruled}
\newfloat{listing}{tb}{lst}{}
\floatname{listing}{Listing}
\title{Neighbor Auto-Grouping Graph Neural Networks for Handover Parameter Configuration in Cellular Network}

\author {
    Mehrtash Mehrabi\textsuperscript{\rm 1,\rm 2},
    Walid Masoudimansour\textsuperscript{\rm 1},
    Yingxue Zhang\textsuperscript{\rm 1},
    Jie Chuai\textsuperscript{\rm 1},
    Zhitang Chen\textsuperscript{\rm 1},
    Mark Coates\textsuperscript{\rm 3},
    Jianye Hao\textsuperscript{\rm 1, 4},
    Yanhui Gen\textsuperscript{\rm 1}
}
\affiliations {
    \textsuperscript{\rm 1} Huawei Noah's Ark Lab, 
    \textsuperscript{\rm 2} University of Alberta,
    \textsuperscript{\rm 3} McGill University, 
    \textsuperscript{\rm 4} Tianjin University \\
    \texttt{\{mehrtash.mehrabi, walid.masoudimansour, yingxue.zhang, chuaijie, chenzhitang2, haojianye, geng.yanhui\}@huawei.com, mark.coates@mcgill.ca}
}

\usepackage{bibentry}

\begin{document}

\maketitle

\begin{abstract}
The mobile communication enabled by cellular networks is the one of the main foundations of our modern society. Optimizing the performance of cellular networks and providing massive connectivity with improved coverage and user experience has a considerable social and economic impact on our daily life. This performance relies heavily on the configuration of the network parameters. However, with the massive increase in both the size and complexity of cellular networks, network management, especially parameter configuration, is becoming complicated. The current practice, which relies largely on experts' prior knowledge, is not adequate and will require lots of domain experts and high maintenance costs. In this work, we propose a learning-based framework for handover parameter configuration. The key challenge, in this case, is to tackle the complicated dependencies between neighboring cells and jointly optimize the whole network. Our framework addresses this challenge in two ways. First, we introduce a novel approach to imitate how the network responds to different network states and parameter values, called auto-grouping graph convolutional network (AG-GCN). During the parameter configuration stage, instead of solving the global optimization problem, we design a local multi-objective optimization strategy where each cell considers several local performance metrics to balance its own performance and its neighbors. We evaluate our proposed algorithm via a simulator constructed using real network data. We demonstrate that the handover parameters our model can find, achieve better average network throughput compared to those recommended by experts as well as alternative baselines, which can bring better network quality and stability. It has the potential to massively reduce costs arising from human expert intervention and maintenance.
\end{abstract}

\section{Introduction}\label{sec:introduction}
The rapid growth in the number of devices that need real time, high quality connection to the internet (e.g., internet of things (IoT) devices, health monitoring equipment, devices used for online education and remote working, autonomous vehicles, etc.) makes it essential to improve cellular network performance. Unsatisfactory user experience and network interruption have negative impacts in our modern society. Thus, improving the cellular network has both economic and social impact towards achieving United Nations Sustainable Development Goals (UNSDGs) \cite{weisenborn2018united, 5gimpact}.
Moreover, it can highly contribute to enhancing infrastructure, promoting sustainable industrialization, fostering innovation, responsible consumption, enabling sustainable cities and communities, and promoting decent work and economic growth \cite{gohar2021role, rao2018impact, siriwardhana2021role}.

The performance of a cellular network relies heavily on its parameter configurations and it is becoming more crucial, as the number of mobile users continues to grow rapidly \cite{mobile_users}. 
These parameters govern access control, handover, and resource management \cite{dahlman20134g, cellularOpt}. 
One of the factors that has a significant impact on the quality of service (QoS) in such networks is the handover parameter \cite{tekinay1991handover}.
We provide more details concerning this parameter and its effects in the supplementary materials, Sec.~\ref{sec:handover}.

Optimizing handover parameters is one of the most common approaches to guarantee minimum service delay or interruption and improve coverage and throughput \cite{mu2014conflict}. However, with the massive increase in both the size and complexity of cellular networks, parameter configuration is becoming  complicated. The current practice, which relies largely on experts' prior knowledge, is inadequate, requiring many domain experts and leading to high maintenance costs.

One of the key challenges in the network parameter optimization problem is the complex spatial and temporal dependencies in the cellular network. Any employed algorithm should be capable of tracking the non-stationary changes in the environment, i.e., the fluctuations of user number, network load, etc. \cite{agiwal2016next}. 
Also, due to the diverse characteristics of cells across the network, the best parameter configuration for one cell may not be optimal for another and parameter configuration of one cell not only affects its own performance, but also affects its neighbors' \cite{dahlman20134g}.
Therefore, there are strong interactions between neighboring cells which become extremely complicated in heterogeneous network.
Consequently, developing an algorithm that can adapt to the temporal dynamics and cell diversity in real networks is essential for parameter configuration \cite{jiang2016machine}.

The current cellular network deployments are highly dependent on human designed rules or analytical models based on domain knowledge and assumptions about the network dynamics which is far from optimal.
They only consider a limited number of network states (e.g., user distribution, channel quality, etc.) and parameters, and cannot capture the complex relationships between network states, parameter configurations and network performance.
Also, the assumptions of the network dynamics, based on which the rules/models are developed, are often simplified without considering the non-stationary changes in real environments, which degrades their performance. 
Finally, these rules/models may not be able to deal with the cell diversities in the network which makes them sub-optimal \cite{challengesOpt}.

Recently, data-driven approaches based on machine learning (ML) have been extensively used for parameter configuration and network management in cellular networks \cite{yuliu2017, tabfar2012, riihijarvi2018machine, infocom_dlCell, dynamicprogramming}.
It has been shown that the multi-layer perceptron (MLP) can be considered as a universal function approximator \cite{goodfellow2016deep}.
Thus, in environments such as cellular networks where there is lack of an accurate analytical model and the network is highly dynamic, neural-network-based methods can be used to achieve high-accuracy prediction.
ML models can utilize high dimensional information and approximate complex functions to fully describe the relationship between network states, parameter configurations and network performance metrics, which cannot be achieved by human experience. 

In order to address the above challenges, we investigate two important questions: 1) \textit{\textbf{Modeling}}: how to model the spatial and temporal dependencies of the cellular network? 2) \textit{\textbf{Decision-making}}: how to choose the parameter values to jointly optimize the overall performance of interconnected and interacting cells?
We first, propose a ML-based model to precisely imitate the cellular network environment and then, use it to configure the parameters. 

We demonstrate that the handover parameters recommended by our model can achieve better average network throughput compared to the existing methods and our approach can massively reduce costs from human expert intervention and maintenance.
It opens up the potential for high-quality internet access to geographical areas that are currently under-served by the cellular network. Besides, this framework can bring new possibilities for important applications to under-developed regions including online education, health monitoring devices by improving their real time connection~\cite{attaran2021impact}.

Our main contributions are summarized as follows. 
\begin{itemize}
    \item We propose a novel method to model the impact from the neighbors of each cell in a distinguishable way to capture the complex spatial dependencies of the network.
    \item We consider  the changing dynamics of the network in our reward model to better reflect the temporal dependencies.
    \item We introduce a multi-objective optimization strategy based on the model to consider several performance metrics and improve the overall network throughput, which has the potential for high social impact applications.
\end{itemize}

\section{Background and Related Work}\label{sec:related_work}
The adjustment of handover parameters helps to balance the traffic load in the network and it can dramatically affect the network throughput. During the handover process in cellular networks, in order to guarantee an acceptable service quality, a user equipment (UE) must monitor the reference signal received power (RSRP) of the serving cell \cite{ltestd}. As soon as the RSRP drops below a pre-defined threshold (called A2-threshold), the UE starts to report measurements to its serving cell and prepares for handover. Increasing the value of A2-threshold decreases the number of UEs in the serving cell in which the handover is triggered, and this spreads the serving cell's load to its neighbors, resulting in a significant change of throughput for the serving cell and its neighbors. While improving the load balance of the network, this can have adverse effects on the network performance since it forces frequent handovers which requires a considerable amount of bandwidth for measurement reporting and causes a drop in network throughput. Decreasing the value of A2-threshold, on the other hand, may cause a poor experience for edge UEs and lead to repeated connection loss due to weak signal. In attempt to solve the problem of optimization of the parameters of a wireless network, different techniques such as fuzzy systems, deep reinforcement learning (DRL), and contextual bandit have been used in the literature. 
(see Sec.~\ref{sec:relatedworkRL} for some details).

The use of graph convolutoinal networks (GCNs) \cite{hamilton2017, kipf2017, fan2019graph} has also yielded significantly well-designed models to predict the network traffic and optimize the corresponding parameters. For example, in \cite{zhazha2020}, the authors introduce a novel handover strategy based on GCNs. The handover process is modeled as a directed graph by which the user tries to predict its future signal strength. Other works such as \cite{zhajia2020} introduce novel methods of network traffic prediction combined with a greedy search or action configuration method to optimize handover parameters. 
However, these works fail to consider the heterogeneous aspect of the cellular networks.

Despite being effective, none of the above-mentioned methods uses the capacity of the neighbors' information to fully tailor the model to adapt to the spatial characteristics of a cellular network, where the interaction is complex and the network is heterogeneous. Also, despite the fact that these techniques consider some important measures of optimization, none of them approaches the problem at hand by considering two of the most important measures simultaneously (especially from the users' perspective): load balancing and throughput.
In this article, we propose an effective and efficient framework that models the network as a heterogeneous graph where we learn an implicit interaction type for each neighboring cell. Then, it incorporates the impact of neighboring cells from each interaction group in a unique way. Moreover, in contrast to the available methods in the literature, we exploit two important measures in the network simultaneously, to configure the parameters effectively: throughput and load balancing, which are directly related to the user experience in the network.

\section{Problem Formulation}\label{sec:problemFormulate}
Let us consider a network with $N$ cells, and form $N$ clusters each composed of one of the network cells as its center cell along with its neighboring cells. 
As an example, we choose the optimization of the A2-threshold to investigate the performance of our algorithm.
According to the 3GPP standard \cite{ltestd}, an A2 event is triggered when the received power at user $u$ from cell $n$, $P^{u,n}$, satisfies
\begin{equation}\label{eq:A2_event}
    P^{u,n} + H_{ys} < Thresh, 
\end{equation}
where $H_{ys}$ is the hysteresis parameter to avoid frequent handovers and $Thresh$ is the A2-threshold we are optimizing.


We consider an online optimization process. 
In real practice, network operators are often conservative and only allow a limited number of experiments. 
During the optimization period of $L$ days, and the A2-threshold can be adjusted once for each cell at the beginning of each day. 
For day $t$, let $D_t$ be the total bits transmitted by all the cells, and $T_t$ be the total transmission time. 
We would like to maximize the accumulated network throughput of the optimization period, i.e., $\max \sum_{t=1}^L \frac{D_t}{T_t}$.
Maximizing the overall network throughput by jointly optimizing the A2-threshold of all cells is difficult. 
The problem becomes even more complicated as the network size increases, which makes a centralized solution not scalable. 
The adjustment of the A2-threshold of one cell only affects its local neighborhood and thus, we convert the centralized problem into a local decision problem. 
That is, each cell only examines its local performance metrics and chooses its own parameter configuration value.

The adjustment of the A2-threshold affects the network throughput via two means: better resource utilization by load balancing, and improved cell throughput with less connection loss and measurement reporting. Consequently, in order to configure it, these two metrics must be considered in the local decision problem. 
The throughput of cell $i$ on day $t$ is highly dependent on its A2-threshold, formulated as $a_t^i$, denoted as $\alpha_t^{i}(a_t^i)$. The load balancing factor in the $i$-th cluster with center cell $i$ on day $t$ with $a_t^i$ is defined as the ratio of the center cell throughput to the average throughput of its neighboring cells, denoted by $\beta_t^{i}(a_t^i)$ and formulated as $\beta_t^{i}(a_t^i) = {\alpha_t^{i}(a_t^i)}/{\bar{\alpha}_t^{i}}$,
where $\bar{\alpha}_t^{i}$ is the average throughput of the neighbors of cell $i$ with action $a_t^i$ and, denoting by $\mathcal{N}_t(i)$ the set of all neighbors of cell $i$ on day $t$, it can be formulated as $\bar{\alpha}_t^{i} = \frac{1}{\vert \mathcal{N}_t(i) \vert}\sum_{j \in \mathcal{N}_t(i)}\alpha_t^{j}(a_t^j)$.
The throughput ratio (rather than traffic/user ratio) is used since different cells have different capacities. This value approaches $1$ when loads of different cells match their capacities.

Our goal is to maximize the overall network throughput by optimizing the two important network performance metrics, namely, throughput ratio $\beta_t^{i}(a_t^i)$ and cell throughput $\alpha_t^{i}(a_t^i)$ for each cell $i\in[1,N_t]$, where $N_t$ is the total number of cells on day $t$, at the same time.
Therefore, we propose the following optimization problem for tuning the A2-threshold for cell $i$:
\begin{equation}\label{eq:opt}
    \argmax_{a_t^i\in\mathcal A}\Big(-\sqrt{\big|1-\beta_t^{i}(a_t^i)\big|}, \alpha_t^{i}(a_t^i)\Big),
\end{equation}
where $\mathcal A$ is the set of all possible values for the A2-threshold in the cellular network.

The challenge of solving the above problem lies in several folds. \textit{First}, since the network performance function is complex, dynamic and unknown, obtaining accurate $\beta_t^{i}(a_t^i)$  and $\alpha_t^{i}(a_t^i)$ is difficult. Instead, in this work, we adopt a data-driven approach to learn reward models and estimate the performance metrics. \textit{Second}, in real-world cases, only a limited experimental budget is allowed by network operators leading to insufficient diverse historical data (state, action pairs) to train a data-driven learning model. In our design, we use a data augmentation technique in the form of neighbor cell augmentation to enrich the features from each cell. \textit{Third}, the handover parameter configuration is affected by adjacent cells. Thus, it is essential to model the information coming from the adjacent cells to achieve accurate reward modeling. 
\textit{Lastly}, optimizing one performance metric greedily might hinder another, thus, how to jointly optimize different performance metrics needs careful consideration. 

\section{Temporal Auto-Grouping GCN for Reward Modeling}\vspace{-.1cm} \label{sec:mlpmodel}

In order to better capture the dependency between each cell and its neighboring cells, we \textit{first} introduce our novel method for neighboring cell feature aggregation. \textit{Second}, we propose a temporal feature aggregation step with recurrent neural networks (RNN) to model the temporal correlation from the historical sequence of the network states. \textit{Third}, we elaborate the overall training process, considering the impact from the neighboring cells, the temporal correlation in the network and the action we aim to optimize. 

\subsection{Spatial Feature Modeling}

The handover parameters heavily impact the learning problem on the graph of the center cells as well as the neighboring cells, hence, we aim to capture the neighboring cells information during our modeling process. 
Recently, message-passing neural networks (MPNNs) in the form of graph neural networks (GNNs) have been introduced and showed to be effective in modeling real world applications with structural information. 
The dependencies in the dataset are modeled using a graph \cite{hamilton2017, ying2018,NGCF_wang19}. In each layer of a GNN, each node's representation includes the features from itself as well as the features from its neighboring nodes (messages sent from the neighborhood). We believe the GNN framework is suitable for handling the dependencies between the center cell and the neighboring cells in cellular networks.  
We present more details on GNN and recent works on homogeneous and heterogeneous graphs in the Sec. \ref{sec:GNNCellular}.


\subsubsection{Graph-Based Cellular Network Modeling}

We construct a graph ${\mathcal G}_t\!\!=\!\!(\mathcal{V}_t,\mathcal{E}_t, \mathbf{X}_t)$ for day $t$, where each node $v\!\!\in\!\! \mathcal{V}_t$ represents one cell and is associated with a feature vector $\textbf{x}_t^{v}\!\!\in\!\!\mathbb{R}^d$ ($v$-th column of $\mathbf{X}_t\!\!\in\!\!\mathbb{R}^{d\times |\mathcal{V}_t|}$), including the statistical properties of node $v$ measured on day $t$. The statistical properties could include several features such as the antenna transmission power, physical resource block (PRB) usage ratio, the amount of data traffic, and the transmission bandwidth. 
These features serve as the node attributes. The edge set $\mathcal{E}_t$ encodes the interactions between cells based on the handover events between pairs of cells. Based on historical data, if any pair of cells has an average number of handover events above a threshold $\tau$, we assume an edge between those two cells. The neighboring set for node $v$ is denoted as $\mathcal{N}_t^g(v)\!\!=\!\!\{u|u\in \mathcal{V}_t,(u,v)\in \mathcal{E}_t\}$. 

Due to the heterogeneous nature of the cellular network, the relationships between the neighboring cells can be complex. Concretely, there might be an implicit $M$ latent relationship types $\mathcal{R} = \{r_1,r_2, \cdots ,r_M\}$ that can be learned to better handle the complex interactions in the cellular networks. Assuming each cell is represented by its states such as PRB usage, traffic, etc. in the network graph, we aim at dividing the neighboring cells into different groups, each of which will provide some information that is shared between the neighbors in that group and help to better capture the rich information from neighboring cells in a distinguishable way. 
Thus, inspired by the above motivation and a recent work~\cite{pei2019geom}, we propose a novel GCN approach called auto-grouping GCN (AG-GCN) to characterize this special property of cellular networks when handling the interactions between neighboring cells. In the following, we elaborate upon the detailed steps to realize our design.

\subsubsection{Neighborhood Augmentation} \label{sec:neighbor_aug}
In cellular network modeling, since the experiment budget is limited, the historical data (state-action pairs) is not diverse enough to train our data-driven model. Besides, since we construct the graph based on the handover events, there are cells that have a very limited number of neighboring cells. Thus, in our design, we use a data augmentation technique in the form of neighbor cell augmentation based on the similarity between cells in a latent space, to enrich the features of each cell. 

\begin{figure*}[t!]
    \vspace{-.3cm}
	\centering
	\includegraphics[width=.93\textwidth]{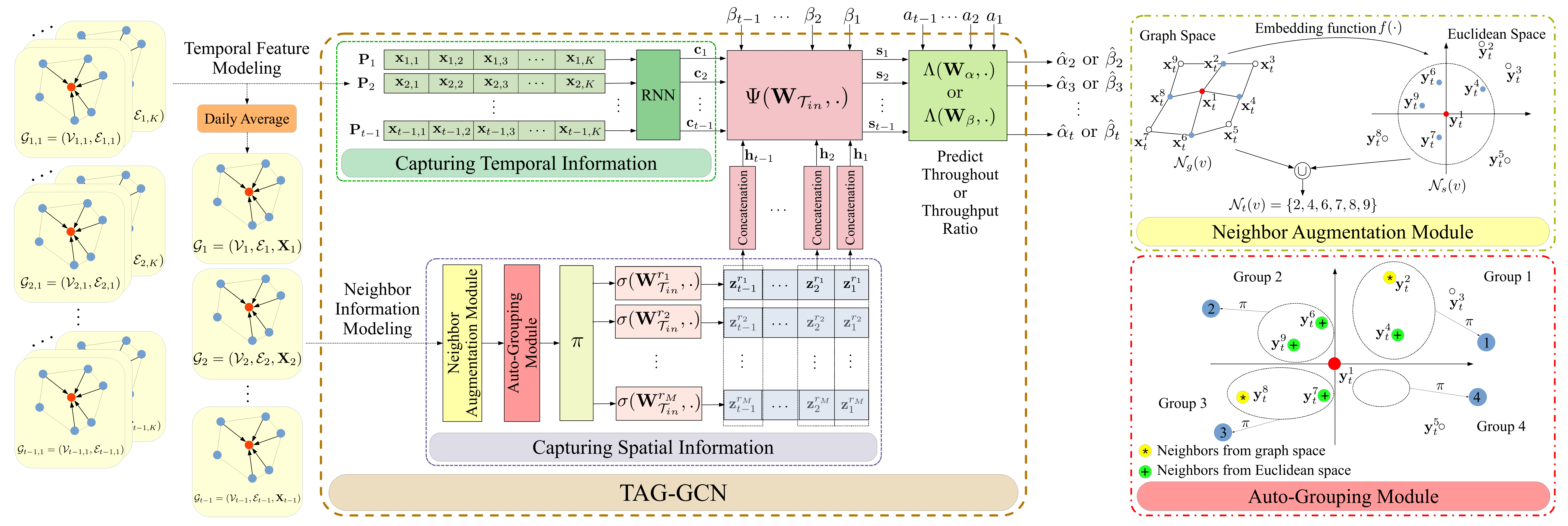}
	\vspace{-.2cm}
	\caption{The flow of information from graph structure to final prediction, used to form the training pipeline of two models for predicting the throughput $\hat{\alpha}$ and the throughput ratio $\hat{\beta}$ for $\mathcal{T}_{in}=\{1,2,\cdots,t-1\}$. 
    In this demo example, the auto-grouping module constructs $M=4$ groups of neighbors, where $l=2$. 
	Empty groups are filled with the average of other groups.}
	\label{fig:nei_aug_aut_grp}
	\vspace{-.45cm}
\end{figure*}

We define a feature transformation function $f(\cdot) : \mathbb{R}^d \rightarrow \mathbb{R}^l$ which maps the input node feature  $\textbf{x}_t^v \in\mathbb{R}^d$ to a latent space $\textbf{y}_t^{v} = f(\textbf{x}_t^{v})\in\mathbb{R}^l$.
In order to capture the long-range dependencies and similarity in the cellular network, we design an additional neighborhood in the latent representation space based on Euclidean distance. For each node $v\in \mathcal{V}_t$, we form the augmented neighborhood ${\mathcal N}_t(v)=\mathcal{N}_t^g(v)\cup \mathcal{N}_t^s(v)$, where $\mathcal{N}_t^g(v)$ and $\mathcal{N}_t^s(v)$ are the neighbors of node $v$ in the original graph and in the latent space, respectively.
The neighbors in the latent space are selected based on their Euclidean distance to the center cell. The $n$ nearest nodes in the latent space are selected to create $\mathcal{N}_t^s(v)$ for cell $v$, where the number of nodes we select based on the feature similarity is equal to the neighborhood size in the original graph $|\mathcal{N}_t^g(v)|\!\!=\!\!|\mathcal{N}_t^s(v)|\!\!=\!\!n$. The neighbor augmentation module in Fig. \ref{fig:nei_aug_aut_grp} illustrates this process.


\subsubsection{Neighborhood Auto-Grouping}
Once we have obtained the augmented neighborhood set, the neighbors in the augmented neighborhood ${\mathcal N}_t(v)$ are divided into different groups by a geometric operator $\gamma$.
Consider node $v$ and its neighbor node $u\in{\mathcal N}_t(v)$. The relation between them on day $t$ is denoted as $\gamma(\textbf{y}_t^{v}, \textbf{y}_t^{u}) : (\mathbb{R}^l, \mathbb{R}^l) \rightarrow \mathcal{R}=\{r_1,r_2,\cdots,r_M\}$. This grouping aims at combining neighbors' information in groups with similar inter-group features. For each group $r_i\in\mathcal{R}$, the neighborhood feature set on day $t$ is defined as $\mathcal{N}_t^{r_i}(v) = \{u|u\in{\mathcal N}_t(v), \gamma(\textbf{y}_t^v,\textbf{y}_t^u)=r_i\}.$
The auto-grouping module in Fig. \ref{fig:nei_aug_aut_grp} demonstrates this process. Note that yellow neighbors (marked with *) are the projected counterparts of the neighbors in the graph space, while the green neighbors (marked with +) correspond to the augmented neighbors from the latent space.

\subsubsection{Conditional Message Passing}
Since the order within each neighbor group should not impact the output of the representation, we apply a permutation invariant function $\pi(\cdot)$ on the neighbors within each group (mean pooling across each feature dimension) and aggregate them separately.
Fig.~\ref{fig:nei_aug_aut_grp} shows an example of the AG-GCN, where $l=2$ and $|\mathcal{R}|=4$ and the representation after the permutation invariant function $\pi(\cdot)$ is shown by black dashed arrows ended to nodes 1, 2, 3, and 4.
Then for each group $r_i\in\mathcal{R}$, a  non-linear transform is further applied as:
\vspace{-0.15cm}
\begin{equation}
    \mathbf{z}_t^{v,r_i} = \sigma\Big(\textbf{W}_t^{v,r_i}\cdot\pi\big(\{\textbf{x}_t^u|u\in  \mathcal{N}_t^{r_i}(v)\}\big)\Big), \vspace{-0.15cm}
\end{equation}
where $\textbf{W}_t^{v,r_i}$ is a learnable weight matrix for the neighbors in group $r_i$ of node $v$ on day $t$, and $\sigma(\cdot)$ is a non-linear function, e.g., $\tanh$. 
Then for each node we aim to aggregate the transformed neighborhood features from their different groups of neighbors in a distinguishable way. The vectors $\textbf{z}_t^{v,r_i}$ for $r_i \in \mathcal{R}$ are further aggregated as $\textbf{h}_t^v = [\textbf{z}_t^{v,r_1}; \cdots; \textbf{z}_t^{v,r_M} ],$ where $[\,\,;\,]$ represents concatenation.
\subsection{Temporal Feature Modeling}
Capturing the trend in the evolution of the states of each cell within a day properly can benefit the prediction of the performance metric for the following day.  We propose to use additional temporal features for each center cell to extract the changing dynamic pattern of its states within each day to further improve the reward model performance.
We assume the samples of the center cell $v$ on day $t$ can be divided into $K$ groups by their temporal order. For all the samples in each group $k$, we take the average network state for each group of samples and denote it as $\textbf{x}_{t,k}^{v}$. We use an RNN layer to capture this temporal dependency of the features from different groups by feeding all the network states as an input sequence $\textbf{P}_t^{v} = \big[{\textbf{x}_{t,1}^{v}}^T; {\textbf{x}_{t,2}^{v}}^T; \cdots; {\textbf{x}_{t,K}^{v}}^T\big]^T\in\mathbb{R}^{K\times d}$, to obtain
\vspace{-.15cm}
\begin{equation}
    \textbf{c}_t^{v} = {\rm RNN}(\textbf{P}_t^{v}, \delta) \in \mathbb{R}^{d'},\vspace{-.15cm}
\end{equation}
where $\delta$ and $d'$ are the set of trainable parameters and the output dimension of the RNN layer, respectively.

\subsection{Overall Training Pipeline}
The main purpose of the model is to estimate the real network‘s response, and predict the throughput ratio and throughput of the center cell for the next day based on the observed network states in the current day. These performance metrics are not only affected by the current day's states, but also highly correlated with the action we choose to configure for the next day.
Thus, we also consider the actions of the next day. 
Furthermore, the throughput ratio and throughput of the next day are highly dependent on the previous performance metrics.
Hence, we consider the current throughput ratio, i.e., $\beta_t^v$, in the prediction process. 

To make the final prediction, the learned representation of the neighborhood by the AG-GCN aggregation, the temporal features of the center cell, and the throughput ratio of the current day, i.e. $\beta_t^{v}$, are concatenated to form the state vector of cell $v$ as $\mathbf{s}_t^{v} = \Psi(\mathbf{W}_t^v\cdot\big[\beta_t^{v}; \textbf{c}_t^{v}; \textbf{h}_t^{v}\big])$, where $\textbf{W}_t^v$ is a learnable weight matrix for node $v$ on day $t$, and $\Psi(\cdot)$ is a non-linear function, e.g., $\tanh$.
Since the final representation should be sensitive to the chosen input action (of which the decision making process will be elaborated in Sec.~\ref{sec:opt}), the throughput ratio and throughput of the next day for cell $v$ are formulated as the output of a non-linear transformation $\Lambda(\cdot)$ function of state and action:
\begin{equation}
    \hat{\beta}_{t+1}^v = \Lambda \big( \mathbf{W}_{\beta}^v.([\textbf{s}_t^{v}; a_t^v])\big),
\end{equation}
\begin{equation}
    \hat{\alpha}_{t+1}^v = \Lambda \big(\mathbf{W}_{\alpha}^v.([\textbf{s}_t^{v}; a_t^v]) \big),
\end{equation}
where $\mathbf{W}_{\beta}^v$ and $\mathbf{W}_{\alpha}^v$ are trainable matrices of node $v$ for throughput ratio and throughput models, respectively.
The overall flow of data from the graph structure to the final prediction is represented in Fig. \ref{fig:nei_aug_aut_grp}. Note that we train two separate models for predicting the throughput and the throughput ratio simultaneously.

In order to properly use the A2-threshold for the prediction, we use the change in this parameter compared to the previous day as the action $a_t^v = A2_{t+1}^v - A2_t^v$,
where $A2_{t+1}^v$ and $A2_t^v$ are the A2-thresholds for cell $v$ on day $t+1$ and $t$, respectively.
The reason for this design choice has twofold. First, the original action space of A2 is large, but the range of the change of action can be smaller by controlling the adjustment steps, making it easier for the model to learn and conduct the decision making step. Besides, the delta action directly reflects the change in the cell coverage/loads, so they are more sensitive to the performance metrics.
To form the training objective, we consider data of $T+1$ consecutive days and  form the pairs $(t,t+1)$, $t\in\{1,2,\cdots,T\}$, to predict the throughput ratio and throughput of the center cell in day $T+1$, trained by minimizing the following loss functions respectively:
\begin{equation}
    \frac{1}{T}\sum_{t=1}^T\frac{1}{N_t}\sum_{v=1}^{N_t} (\hat{\beta}_{t+1}^v - \beta_{t+1}^v)^2 + \lambda_1||\Theta_1||^2, 
\end{equation}
\begin{equation}
    \frac{1}{T}\sum_{t=1}^T\frac{1}{N_t}\sum_{v=1}^{N_t} (\hat{\alpha}_{t+1}^v - \alpha_{t+1}^v)^2 + \lambda_2||\Theta_2||^2,
\end{equation}
where $\lambda_1$ and $\lambda_2$ are the hyperparameters chosen for regularization. ${
\Theta}_1$ and ${\Theta}_2$ represent all the trainable parameters in the models. 
The trained reward model is now able to mimic the real network and predict both throughput ratio and throughput of each center cell for the coming day and can be used to check the impact of actions towards the performance metrics we are considering.

\section{Action Configuration}\label{sec:opt}
As discussed in the earlier sections the main objectives to consider in the action configuration process are load balancing, identified by the throughput ratio, and the cell throughput. 
Hence, the best action for cell $v$ on day $t$, i.e. $a_t^{v}\in{\mathcal A}$, is the one that optimizes the problem in (\ref{eq:opt}).
In general, when dealing with a multi-objective problem, different objectives are often conflicting, and we may not be able to optimize them simultaneously. {{One common way to tackle this problem is to give different objectives weights and optimize the weighted objective value. However, in our scenario, it is difficult to determine the weights and different clusters may require cluster-specific weights.}} Here we break the problem in (\ref{eq:opt}) into two sub-problems, and solve them sequentially. We first optimize the action with respect to the predicted throughput ratio, i.e., $\hat{\beta}_{t+1}^{v}(a_t^{v})$ for cell $v$ on day $t$, where $a_t^{v}\in{\mathcal A}$, and then optimize the throughput $\hat{\alpha}_{t+1}^{v}(a_t^{v})$.

Specifically, the throughput ratio is optimized and we find the set of best $c$ values for $a_t^{v}$, denoted $\mathcal A_c^{v}$, such that 
\vspace{-.15cm}
\begin{equation}
    \min_{a_t^{v}\in{\mathcal A_c^{v}}}\hspace{-.2cm}-\sqrt{\big|1-\hat{\beta}_{t+1}^{v}(a_t^{v})\big|}\ge\hspace{-.3cm}\max_{a_t^{v}\in{\mathcal A}-{\mathcal A_c^{v}}}\hspace{-.3cm}-\sqrt{\big|1-\hat{\beta}_{t+1}^{v}(a_t^{v})\big|}.\vspace{-.15cm}
\end{equation}
Then, our goal is to achieve the maximum possible throughput for cell $v$ on day $t$ and this is through
\vspace{-0.15cm}
\begin{equation}
    \hat{a}_t^v = \argmax_{a_t^{v}\in\mathcal A_c^{v}}~\hat{\alpha}_{t+1}^{v}(a_t^{v}). \vspace{-0.2cm}
\end{equation}
$\hat{a}_t^v$ is then the final recommended action for cell $v$ on day $t$. This procedure for all the $N_t$ cells of the network on day $t$ is presented in
Algorithm~\ref{alg}.



\begin{table}[t!]
    \centering
    {\fontsize{8}{10}\selectfont
    \begin{tabular}{|c|c|c|}
        \hline
         $\gamma(\textbf{y}_t^{v}, \textbf{y}_t^{u})$ &  $\textbf{y}_t^{v}[0] > \textbf{y}_t^{u}[0]$ & $\textbf{y}_t^{v}[0]\leq \textbf{y}_t^{u}[0]$\\ \hline
         $\textbf{y}_t^{v}[1] \leq \textbf{y}_t^{u}[1]$ & 2 & 1 \\ 
         $\textbf{y}_t^{v}[1] > \textbf{y}_t^{u}[1]$ & 3 & 4 \\ \hline
    \end{tabular}}
    \vspace{-.3cm}
    \caption{The relationship operator $\gamma$}
    \vspace{-.3cm}
    \label{tab:relationshipOperator}
\end{table}

\begin{figure}
\vspace{-.4cm}
\begin{algorithm}[H]
\caption{TAG-GCN for Action Configuration}
\label{alg}
\textbf{Input}: $\textbf{P}_t^v$ and $\textbf{x}_t^u$ where, $u\in\mathcal{N}_t(v)$ $\forall v\in[1,N_t]$ \\
\textbf{Output}: $a_t^{v}$ for $\forall v\in[1,N_t]$
\begin{algorithmic}[1] 
\STATE Let $v = 1$ and $\forall j\in[1,N_t]$ set $\mathcal A_c^{j}=\emptyset$.
\WHILE{$v\le N_t$}
\STATE Feed the TAG-GCN model with $\textbf{P}_t^v$ and $\textbf{x}_t^u$, for all neighbor $u\in\mathcal{N}_t(v)$.
\STATE Freeze all inputs of TAG-GCN except actions and predict the performance metrics $\hat{\alpha}_{t+1}^{v}(\cdot)$ and $\hat{\beta}_{t+1}^{v}(\cdot)$ for the input actions.
\FOR{$|\mathcal A_c^{v}|\le \nu$}\vspace{-.15cm}
\STATE $x = \argmax_{a\in\{\mathcal A-\mathcal A_c^{v}\}}-\sqrt{\big|1-\hat{\beta}_{t+1}^{v}(a)\big|}$\;
\vspace{-.1cm}
  \STATE $\mathcal A_c^{v} = \mathcal A_c^{v} \cup \{x\}$
\ENDFOR
\STATE $a_t^{v} = \argmax_{a\in\mathcal A_c^{v}}\hat{\alpha}_{t+1}^{v}(a)$ \;
\STATE $v = v+1$
\ENDWHILE
\STATE \textbf{return} $a_t^{v}$ for $\forall v\in[1,N_t]$
\end{algorithmic}
\end{algorithm}
\vspace{-.8cm}
\end{figure}

\section{Experimental Results}\label{sec:simulation}
The experiments are conducted on a large-scale cellular network simulator constructed from real-world data which presented in Sec. \ref{sec:simulator}. 
We use principal component analysis (PCA) as the mapping function $f(\cdot)$, as defined in Sec.~\ref{sec:mlpmodel}, for obtaining the latent representation in the AG-GCN step. It transforms the original feature into a 2-dimensional space to perform the neighborhood augmentation and the neighbors group assignment process. 
After this transformation, the relationship operator $\gamma$ for the auto-grouping assigns a group to each subset of points in each quadrant of this two dimensional space presented in Table~\ref{tab:relationshipOperator}. The permutation-invariant function $\pi$ applied on each group of neighbors is average in our experiments.

\subsection{Datasets}
To perform our experiments and evaluate the proposed model, two datasets are used in this study (see Sec. \ref{sec:dataset_samples} for more details):

\paragraph{Dataset-A:} A real metropolitan cellular network containing around 1500 cells sampled hourly and collected from Oct. 17 to Oct. 31, 2019. 
Each data sample contains information such as the cell ID, sample time, configuration of cell parameters, and measurements of the cell states.
	
\paragraph{Dataset-B:} Also a real metropolitan cellular network. The network contains 1459 cells, and the data is collected from Sep. 1 to Sep. 29, 2021. Each data sample contains similar information as above.

\subsection{Reward Model Accuracy Evaluation}


\subsubsection{Dataset Generation}
In order to evaluate the prediction accuracy of our model, we use a simulator to modify Dataset-A with a random policy to diversify our network configuration. 
On each day, the A2-threshold for each cell is randomly selected around the default action -100 dBm within the range $[-105, -95]$. 
This approach provides us a fix data buffer with diverse action dataset to train all models and have a fair comparison of their accuracy. For Dataset-B, there exists a reasonable amount of the diversity in the handover parameter configuration, thus we directly use the raw dataset from the live network to perform the training and evaluation.

\subsubsection{Training Process and Metrics}
As samples generated hourly, we aggregate them within each day as described in Sec.~\ref{sec:mlpmodel}. 
To evaluate the model accuracy in predicting cell throughput and throughput ratio, we train the model with the generated pairs $\{(1,2),\cdots,(t-1,t)\}$ for $t=9$ and 12 days for Dataset-A and B, respectively.
At each day $t>2$, data pairs $\{(1,2),\cdots,(t-2,t-1)\}$ are used
as training and validation sets, and $(t-1,t)$ serves as testing set for evaluation across different models.  
We report the mean square error (MSE) to measure the reward model performance.

\begin{figure}[t!]
    \vspace{-.4cm}
    \subfloat[]{\includegraphics[width=.23\textwidth]{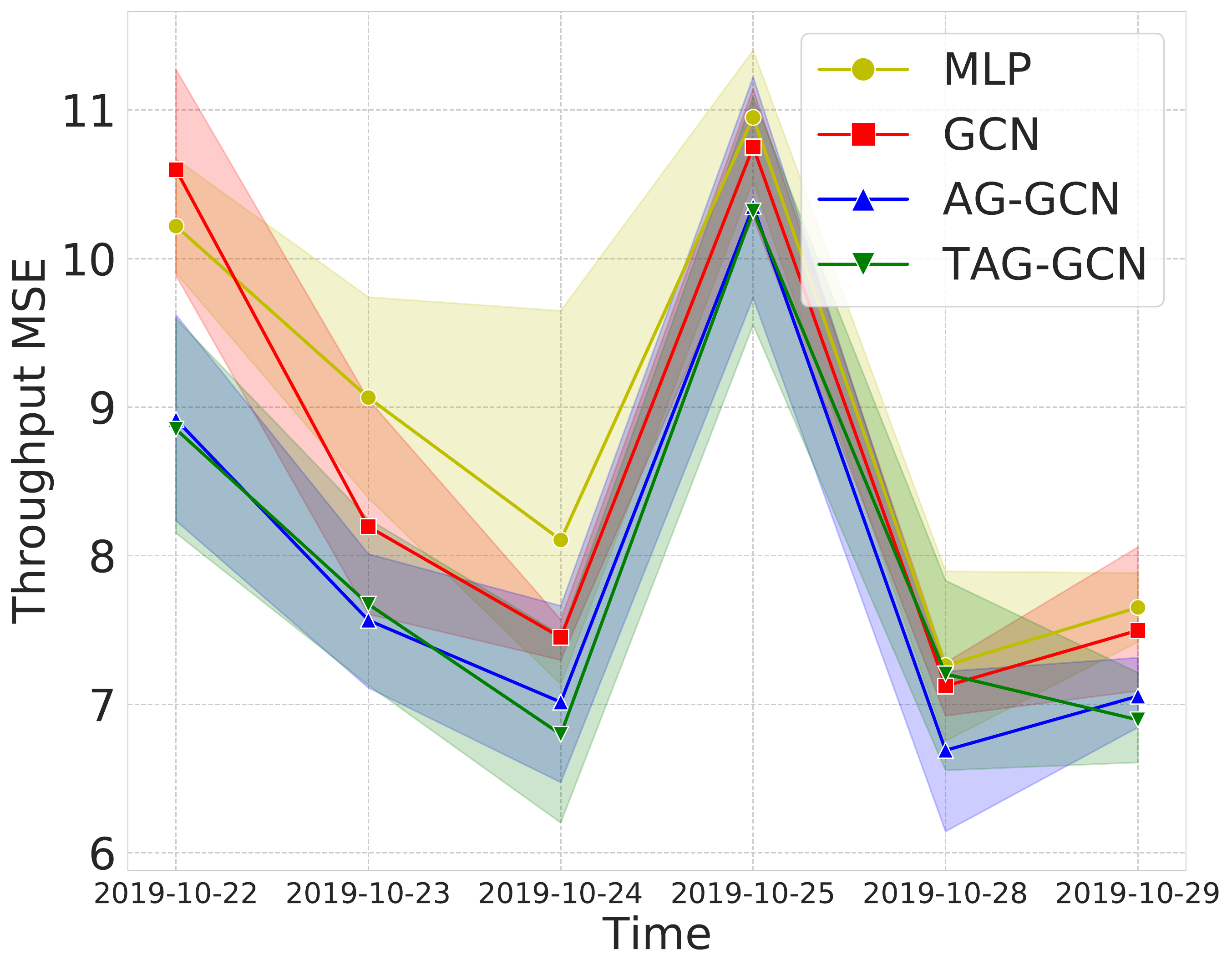}}
	\subfloat[]{\includegraphics[width=.23\textwidth]{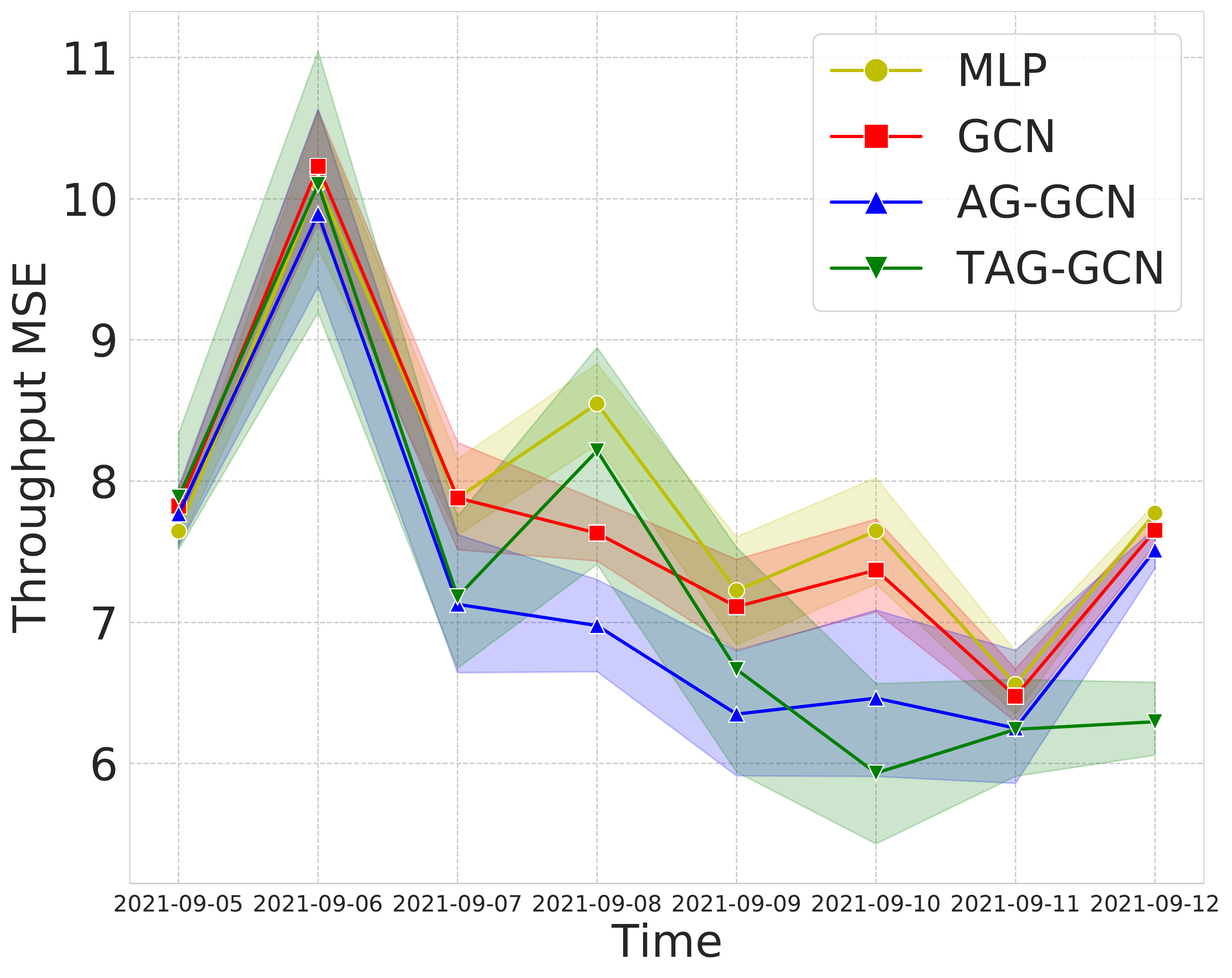}}
	\vspace{-.4cm}
	\caption{Achieved MSE of the throughput for test data of (a) Dataset-A and (b) Dataset-B for different methods}
	\vspace{-.7cm}
	\label{fig:TAG-GCN_test_mse_thrp}
\end{figure}


\subsubsection{Comparison with Benchmark Models}\label{sec:benchmark}

It is important to note that, the social impact of this work has not been addressed by ML approaches the same way as we propose. 
Due to the uniqueness of our problem, existing solutions for optimizing handover parameters are either not appropriate to solve it or there is no apparent way to directly adapt them to our problem. 
For instance, traditional handover optimization methods rely on designing fuzzy rules based on different measures of QoS in the network~\cite{vasmah2012}, however, designing proper rules is complex and cannot handle the change in highly dynamic systems well. Instead, we hope to use a data-driven approach, among which the (deep) RL method gains the most attention ~\cite{caolu2018,wanli2018}. However, in this type of problems, the network provider only allows limited exploration of the parameter values (e.g., allows changing the A2 value once a day) to ensure the stability of the network. Thus, we only have limited days for exploring the best action. RL models, nevertheless, usually need longer episodes to optimize the accumulated return. 
Despite this fact, we made some preliminary attempt, presented in Sec.~\ref{sec:appendix_rl}, to adapt the RL paradigm from the literature into our problem which did not show any advantage over our simpler design.

In order to show the effectiveness of our proposed reward model, we compare it with alternative designs for the prediction model.
It should be mentioned that all of these models are our contribution.
In the Sec.~\ref{sec:benchmark}, we summarize our benchmarks and the properties of each model.
The first model is MLP, where we only use the features of the center cells and ignore the neighboring cells' features.
In GCN model, we follow the typical GCN formulation~\cite{hamilton2017} and process the network as a homogeneous graph where the neighbor information is aggregated jointly without distinction. The AG-GCN model ignores the temporal dependencies of the data which we consider in TAG-GCN model.
In Fig.~\ref{fig:TAG-GCN_test_mse_thrp}, we compare the prediction accuracy of these models for throughput in Dataset-A and B.  
  We observe on average the best accuracy for the test set is achieved by AG-GCN and TAG-GCN, with TAG-GCN performing marginally better on the average rank metric across the evaluation days, indicating that our neighbor aggregation and temporal features extraction have a considerable impact on the reward modeling for cellular networks.
  The same results also achieved for the throughput ratio model.
  The same test is performed for throughput ratio and included in the supplementary materials in Fig.~\ref{fig:TAG-GCN_test_mse_ratio}. 

\begin{figure}[t!]
    \centering
    \includegraphics[width=.4\textwidth]{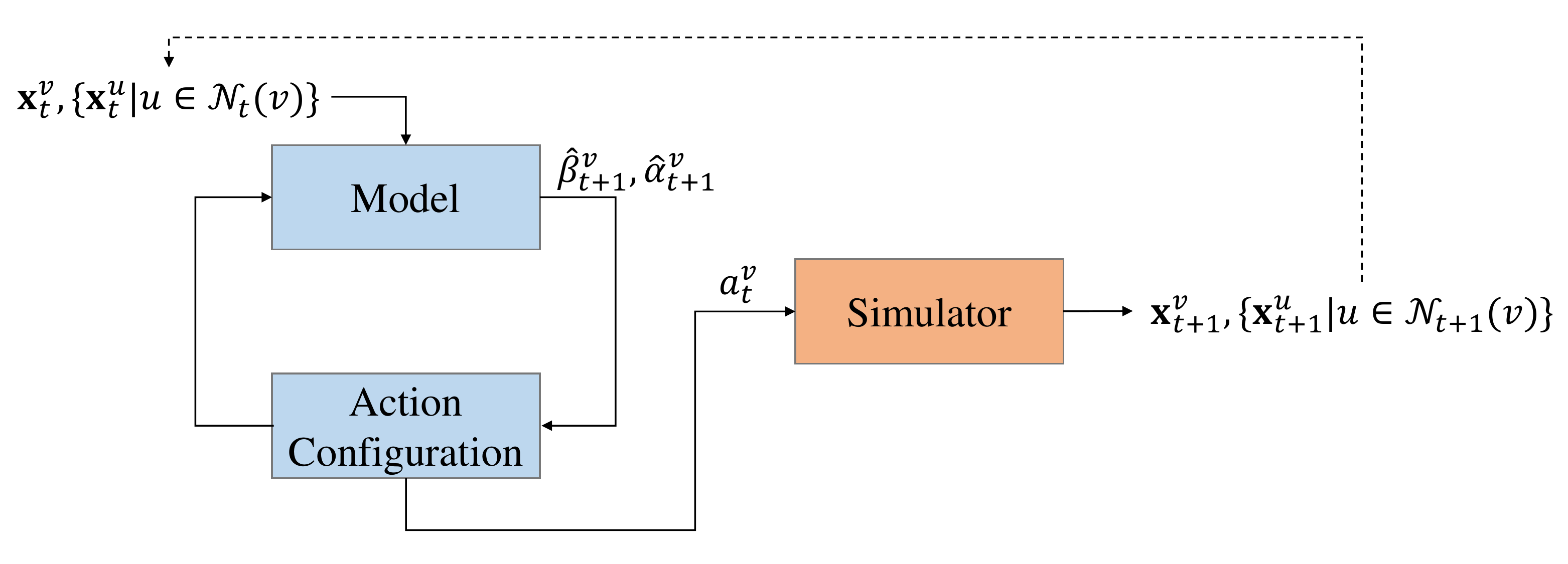}
    \vspace{-.45cm}
    \caption{The process of action recommendation by the trained model and the simulator}
    \vspace{-.5cm}
    \label{fig:training}
\end{figure}

\subsection{Overall Parameter Optimization Performance}
\subsubsection{The Action Recommendation Process}
In the following experiments we use the presented models to recommend the actions for Dataset-A.
The actions in day 1, i.e., Oct. 17, has been set to the default action which is -100~dBm.
Unless otherwise stated, the action for the second day, i.e., Oct. 18, is initialized by a set of random actions around the default action in the range of $[-105, -95]$.
The model is trained iteratively on each day and used to recommend actions for the next day.
The process is depicted in Fig.~\ref{fig:training}, where states of the cells on day $t$ are given to the trained model to predict performance metrics of the network on day $t+1$ and the action $a^v_{t}$ is adjusted for each cell based on the predictions.
Finally, the network states and performance measurements for day $t+1$ are computed according to the new selected action by the cellular network simulator and used for model training and action recommendation in the following day.

\subsubsection{Baseline Performance Bounds}
In addition to the result achieved by the actions recommended by the models, we use three baseline performance bounds achieved by the \textit{default A2-threshold}, the \textit{expert rule}, and the \textit{optimal actions} of the simulator.
As stated before the default A2-threshold value is $-100$~dBm and this is used as the lower bound in the following experiments.
The optimal actions in the simulator are obtained by brute-force search and it introduces the upper performance bound. 
The expert rule-based method provided by experienced network operator is a simple rule presented in the Sec.~\ref{sec:expertrule}.
The performance achieved by the expert rule is better than the default action. We hope to use our proposed learning based framework to further fill the gap with the reward achieved by optimal actions. 



\begin{figure}[t!]
    \centering
    \includegraphics[width=.45\textwidth]{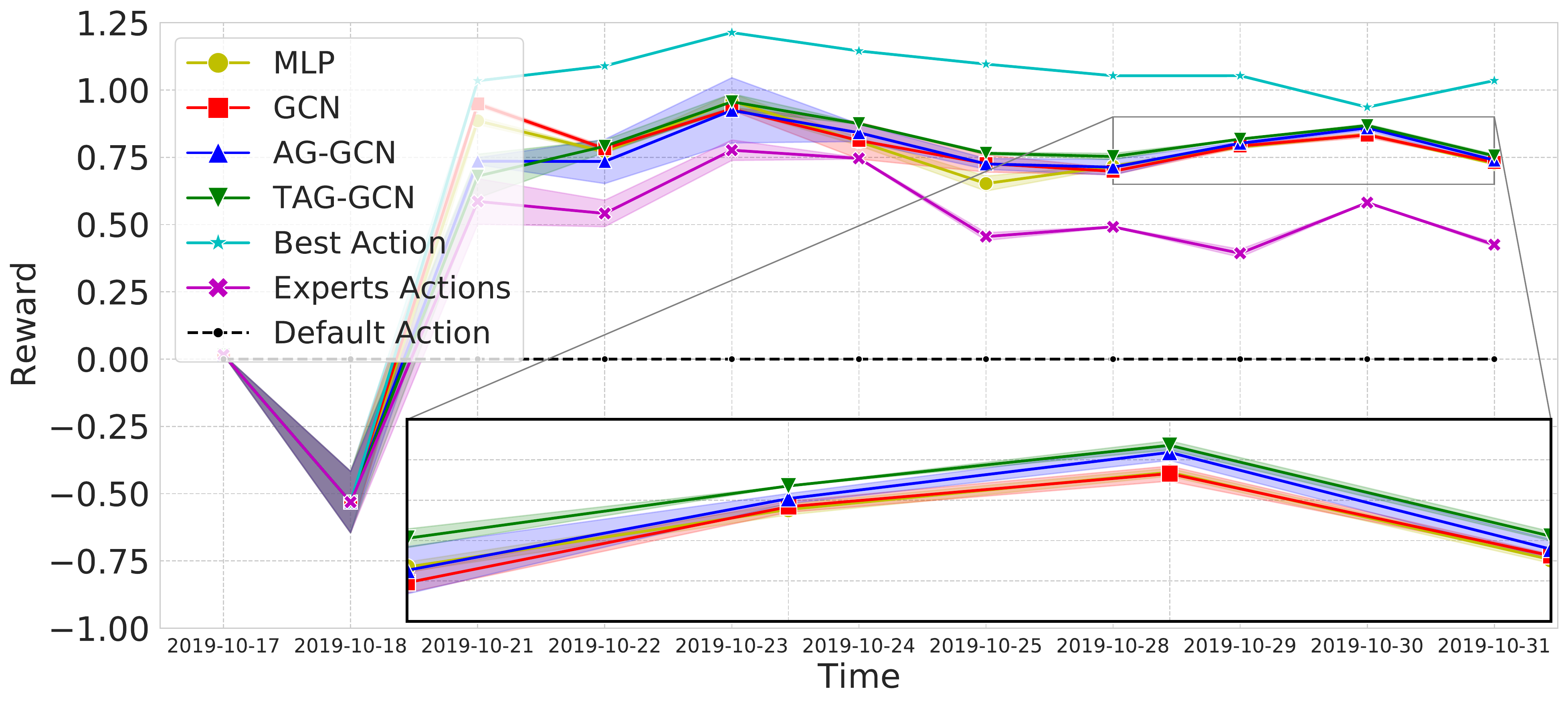}
    \vspace{-.4cm}
    \caption{Performance comparison of different models along with optimal action curve initialized with random actions}
    \vspace{-.5cm}
    \label{fig:ModelPerformance}
\end{figure}

\subsubsection{Results}
\begin{figure}[t!]
	\centering
	\includegraphics[width=.45\textwidth]{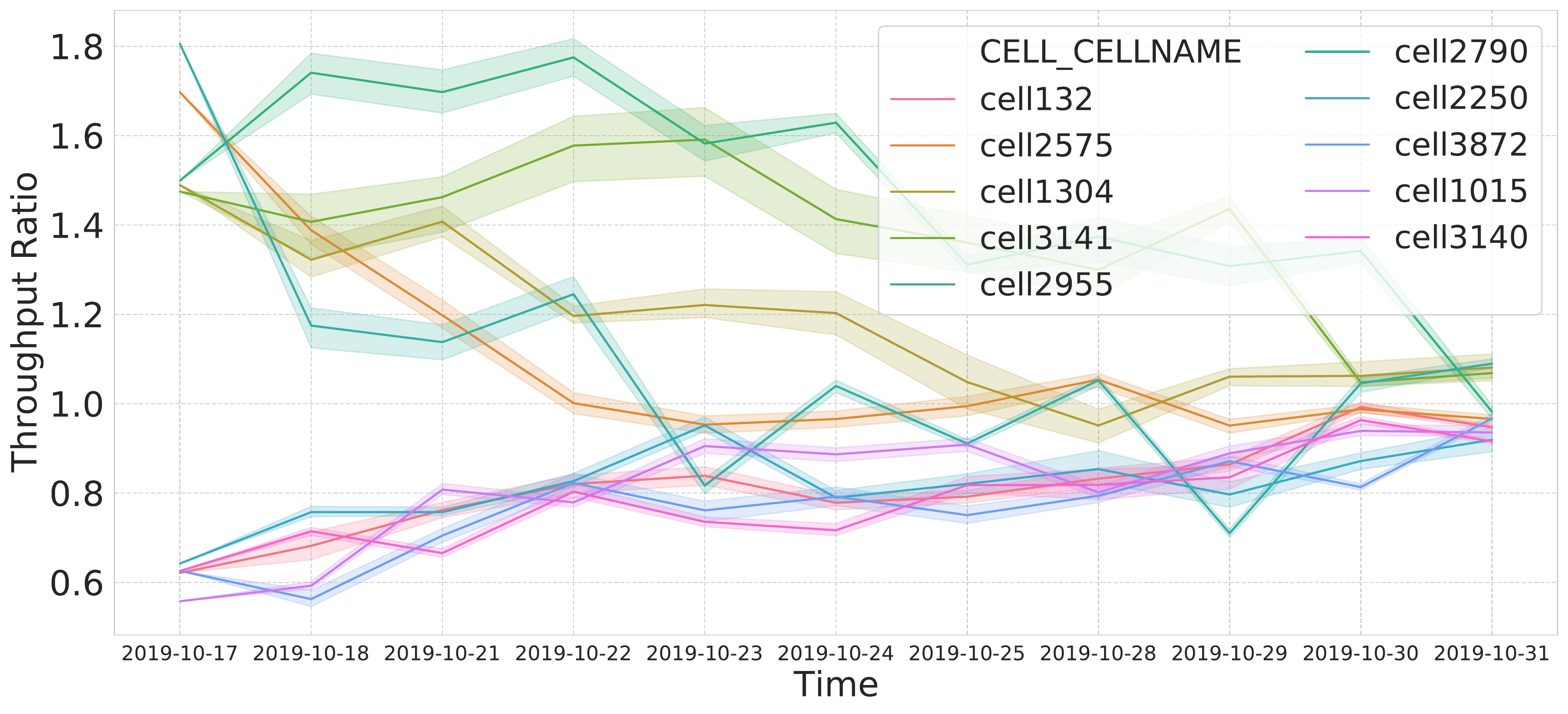}
	\vspace{-.4cm}
	\caption{Trend of the throughput ratio for sample clusters}
	\vspace{-.5cm}
	\label{fig:cell_thrp_ratio_progress}
\end{figure}

In the following experiments, we compare the performance of the network under the actions recommended by different models in terms of the network throughput as defined in Sec.~\ref{sec:problemFormulate}. We plot the trajectory of the throughput difference to the default A2-threshold baseline (dash black line) in Fig.~\ref{fig:ModelPerformance}. 
We repeat all the experiments 20 times for all models, where each run uses the same set of random actions on the first action exploration day (Oct. 18) for all the models.
We also show the performance achieved through the expert rule action recommendation, default action, and the optimal actions of the simulator (random actions are also used on Oct. 18 for the curve of the optimal action). 
The quantitative results for Fig.~\ref{fig:ModelPerformance} are summarized in Table~\ref{tab:num_res_diff_models} in the supplementary materials. 
TAG-GCN can achieve better average throughput in the final days which indicates the importance of our auto-grouping GCN design to tailor the heterogeneous property of the cellular networks.
Besides, as expected, all the learning-based models can beat the expert rule algorithm which is highly dependent on human experience and is unable to recover from the performance degradation due to bad random initialization on the first day. Furthermore, to show the effectiveness of our proposed model in terms of load balancing and enhancing cluster throughput ratio, we illustrate the progress of this ratio achieved by TAG-GCN for some selected severely unbalanced cells in Fig.\ref{fig:cell_thrp_ratio_progress}. As it can be seen, the throughput ratio of the clusters form a trajectory that converges to 1 which is the ideal target value. 
More ablation studies are presented in Sec.~\ref{sec:ablationStudies}.

Based on the above experiments, our proposed ML-based solutions can improve the network performance and optimize the handover process compared to the conventional methods such as using the default action or human experts rule-based methods. Moreover, the automation of the parameter optimization process achieved by our ML-based solutions reduces the domain expert’s intervention and, hence, the management cost of network operators and improves the maintenance efficiency of cellular networks. Consequently, the proposed solutions open up the possibilities to provide reliable and high-quality network access even to geographical areas that are currently underserved by the cellular network. This can bring exciting new opportunities to these regions such as remote education, remote working, health monitoring, video streaming, etc.

\section{Conclusion}\label{sec:conclu}
In this paper, we study the handover parameter configuration problem in cellular networks. 
We propose a reward prediction model to accurately imitate the cellular network and estimate the performance metrics.
Our proposed model, i.e., TAG-GCN, investigates the impact of the adjacent cells and differentiate their impact on the center cell of each cluster. 
We also consider the network changing dynamics in our model to learn the temporal dependencies in the data. Based on the reward model, a novel multi-objective parameter configuration strategy is proposed to perform the optimization for each cluster and balance the performance metrics in each neighborhood.
The conducted simulations shows the superiority of TAG-GCN which has a huge potential social impact by improving the cellular network parameters and providing massive connectivity and high coverage with a balanced traffic across the network.
Hence, this can help the widespread adaptation of new technologies to benefit many sectors such as health and education.

\bibliography{ref}

\newpage

\appendix
\section*{Supplementary Materials} \label{sec:app}
\section{Extended Background and Related Work}
\subsection{Handover in Cellular Networks} \label{sec:handover}
\begin{figure}[t!]
    \centering
    \includegraphics[width=.4\textwidth]{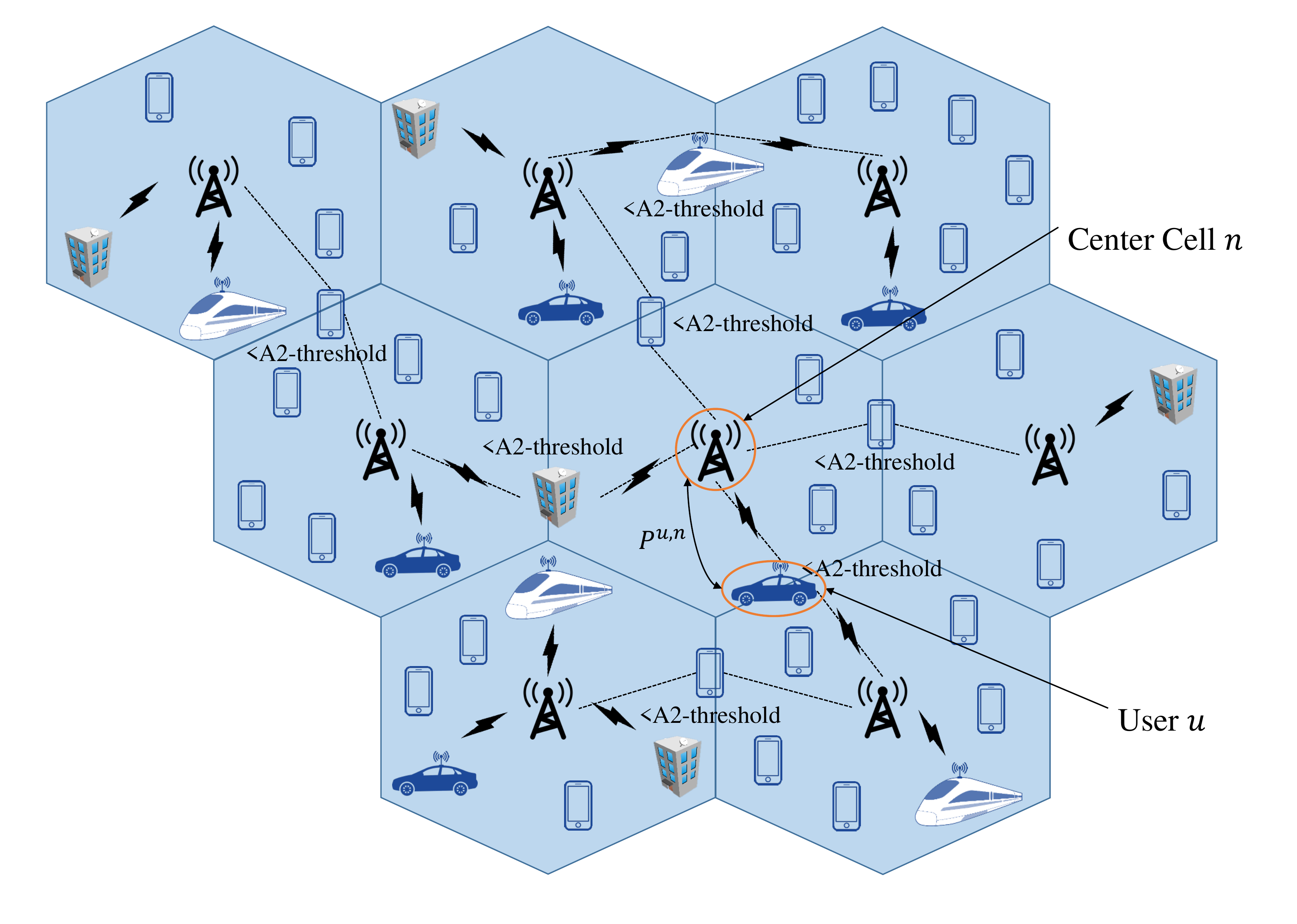}
    \vspace{-.55cm}
    \caption{A typical structure of a cluster in the cellular network serving different types of users.}
    \vspace{-.3cm}
    \label{fig:cells}
\end{figure}

A cellular network is formed by many cells distributed over land areas, each served by at least one fixed-location base station (BS) providing wireless links with a wide geographic area coverage to support many users.
A cluster in a cellular network (consisting of a cell and its neighboring cells) is shown in Fig.~\ref{fig:cells} where, user $u$ is crossing one cell's coverage area and monitors the received power from cell $n$ to check the handover criteria. 
The cell BS is responsible for monitoring the strength of the signals received by served users, which can degrade when a user travels from one cell to another.
The BS should trigger a handover process at the proper moment to avoid any service interruption and transfer the user to another cell BS that is receiving the strongest signals \cite{cellularOpt, tekinay1991handover}.

There are multiple parameters that impact handover in a network. These parameters generally are grouped in three categories. {\em RSRP} and {\em RSRQ} parameters are indicators of the signal strength and quality of the serving station, respectively, {\em Hysteresis} parameters act as a tolerance margin to avoid the Ping Pong (PP) effect, and {\em Time To Trigger (TTT)} parameters  are set such that short term violations of handover conditions are ignored to avoid the PP effect.
Also, different measures such as handover failure (HOF) rate, handover frequency, Ping-Pong (PP) rate, network throughput, and load balancing have been used in the literature to assess the effectiveness of the proposed algorithms for optimizing handover parameters. When performing optimization in a real network, some of these measures may conflict. Therefore, it is crucial to consider a combination of these quantities to find a solution that satisfies different requirements of the network.

\subsection{Related Work}\label{sec:relatedworkRL}
Different methods have been used in the literature for solving the problem of optimization of handover parameters in a wireless network. Fuzzy system handover algorithms, for example, have been used in \cite{vasmah2012}. Such techniques, however, are not scalable despite being accurate and stable. Deep reinforcement learning (DRL) is another technique that has been used to solve the handover optimization problem. In \cite{caolu2018}, the authors propose a framework based on DRL where actions are flexible and can be chosen by the user, and the objective of the optimization is the throughput and the handover count. In \cite{wanli2018}, the authors propose a framework in which the UEs first are clustered based on their usage pattern and then the handover process is optimized in each cluster by a DRL method. Another newly introduced approach to the problem of cellular network parameter optimization is the Contextual Bandit model proposed in \cite{infocom_dlCell}in which the throughput of the cells is used as a performance metric for optimizing some of the cell parameters.

The above mentioned methods, despite being effective, do not exploit the information from the neighboring cells which can be used to develop a more effective model of the network. Moreover, two of the most important measures from the user's standpoint are throughput and load balancing, none of which are considered in the above models simultaneously. Thus, in the current work, we propose a technique which remedies these shortcomings by utilizing the neighbors' information, and optimizing the parameters of the network using two significantly important objectives, namely, load balancing and throughput.

\subsection{Graph Neural Networks for Cellular Network} \label{sec:GNNCellular}
\begin{figure}[t!]
    \centering
    \includegraphics[width=0.5\textwidth]{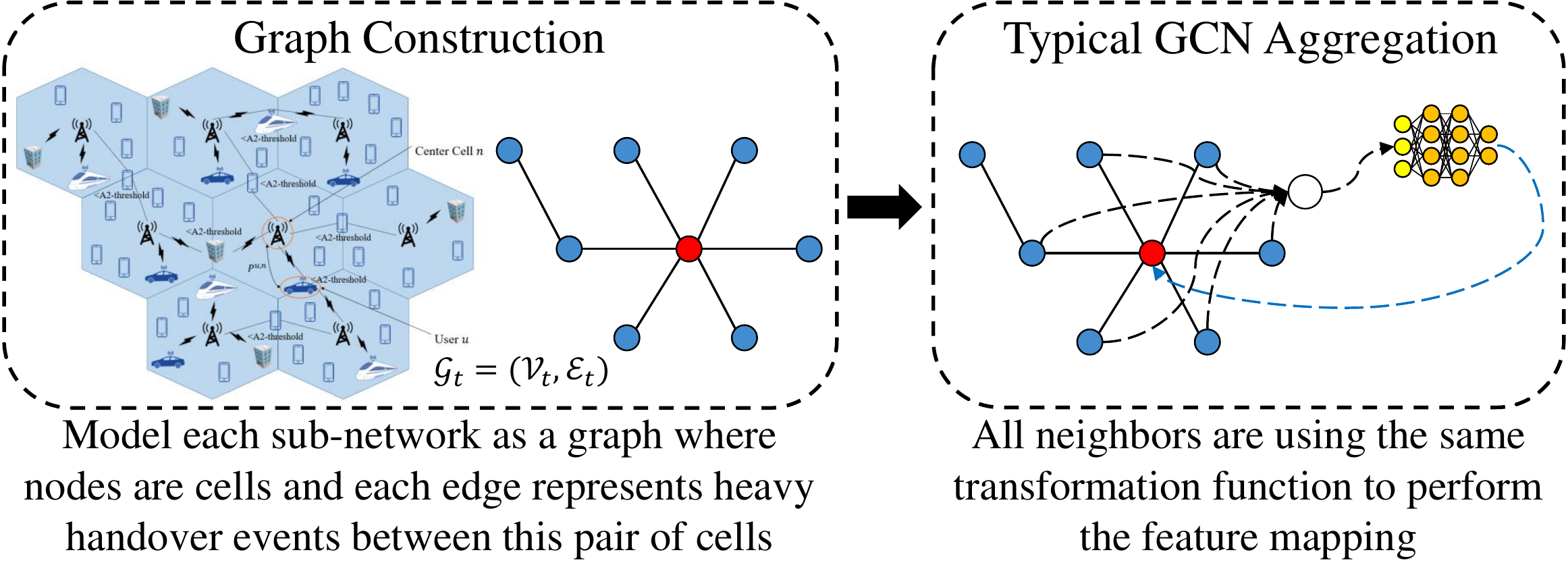}
    \vspace{-.7cm}
    \caption{Commonly used GNN design.}
    \vspace{-.3cm}
    \label{fig:normal_gcn}
\end{figure}

Successful applications of GNNs mostly rely on a homogeneous graph structure. For example, in social networks or recommender systems, each edge represents a positive correlation between the adjacent nodes~\cite{kipf2017, fan2019graph}. The MPNN framework~\cite{gilmer2017neural} applies the same transformation function to every neighbor and aggregates across them to obtain the neighbor representation, shown in Fig.~\ref{fig:normal_gcn}. There is a rich literature that studies how to process a heterogeneous graph with multiple relation types~\cite{schlichtkrull2018modeling, zhang2019heterogeneous}. They are established on the prior knowledge of the given relation type between each pair of nodes and then use a meta-path design to distinguish the information coming from different types of neighbors. Such an approach is not applicable to our problem. 

\section{Experimental Details, Results, and Ablations} \label{sec:app}

\subsection{Simulator Construction}\label{sec:simulator}
The experiments are run through a proprietary cellular network simulator which uses statistical models to simulate the network response to any change in the parameter values. This simulator is designed based on the network corresponding to Dataset-A and uses the states of the network cells along with their action to simulate the throughput and the traffic of the cells.

Dataset-A is collected under the default network configuration and the value of A2-threshold is not changed during the collection period. To evaluate our model and the action configuration strategy, the simulator needs to simulate the network performance given arbitrary A2-threshold values of the cells. More precisely, let $\textbf{x}_t^i$ represents cell $i$'s states at day $t$ under the default A2-threshold configuration, and $A2_t^i$ be the configured A2-threshold value for cell $i$ at day $t$, the simulator will output
\begin{equation}
    \Tilde{\textbf{x}}_t^0, ..., \Tilde{\textbf{x}}_t^{N_t}, r_t^0, ..., r_t^{N_t} = g(\textbf{x}_t^0, ..., \textbf{x}_t^{N_t}, A2_t^0, ...,A2_t^{N_t}),
\end{equation}
where $r^i_{t}$ is cell $i$'s throughput, and $\Tilde{\textbf{x}}_t^{i}$ is the observed states of cell $i$ under the new configurations at $t$.

\begin{figure}[t!]
    \centering
    \includegraphics[width=.45\textwidth]{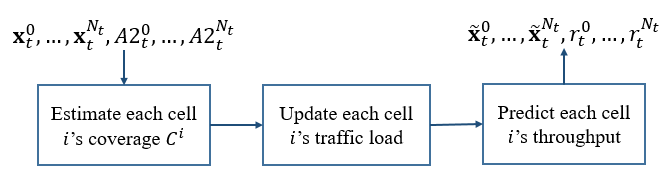}
    \vspace{-.5cm}
    \caption{The internal workflow of the simulator.}
    \vspace{-.3cm}
    \label{fig:simu_flow}
\end{figure}

Fig.~\ref{fig:simu_flow} gives a simple sketch on the internal workflow of the simulator. During simulation, when $A2_t^0,...,A2_t^{N_t}$ are configured at day $t$, the cell states $\textbf{x}_t^0, ..., \textbf{x}_t^{N_t}$ are read from the collected data, and inputted to the simulator. The cell coverage $C^i$ for each $i$ is first computed by a built-in function that considers cell $i$'s frequency, bandwidth and $A2_t^i$. Then cell $i$'s traffic load (including the number of users and the amount of data bits for transmission) at day $t$ is redistributed among itself and its neighbors based on the change of $C^i$ and $C^j$ with $j \in \mathcal{N}_t(i)$. After each cell's traffic load is updated, the throughput $r_t^i$ is predicted based on cell $i$'s traffic load. The load-throughput prediction model for each cell is pre-trained from the collected historical data and stored when the simulator is initialized. The value $r_t^i$ is further adjusted by a factor that considers the throughput loss due to measurement reporting or connection loss. Finally, the simulator outputs $r_t^0,...,r_t^{N_t}$. As each cell's traffic load (part of the cell states) is modified, the updated states $\Tilde{\textbf{x}}_t^0, ..., \Tilde{\textbf{x}}_t^{N_t}$ are also outputted.

\subsection{Dataset Samples} \label{sec:dataset_samples}
Dataset-A and B contain wide range of cell states (e.g., the number of total users within the cell, the number of active users, the cell average CQI, the cell traffic load, etc.) and performance indicators (e.g., the average cell throughput, the edge user throughput, etc.).
The neighbor relation information between cells in both datasets are collected and the hourly average handover counts between neighbor cells are recorded. The weekend dates are excluded from Dataset-A since these days are not considered as rush hour usage for this region of cellular networks.

We present the values of four different features of three random cells in Dataset-A in Tables~\ref{tab:dataset_A}.
Note that the represented values are collected in the first day.
As seen the range of variation across different cells and different hours for each single cell is too high and that is the reason we consider the modules to handle spatial and temporal dependencies in our model to better imitate the real-world environment.
We have almost the same pattern in Dataset-B as well for these presented features.

\begin{table}[]
	\centering
	\resizebox{\columnwidth}{!}{
	\begin{tabular}{|c|c|c|c|c|c|}
	\hline
		& Time &  Cell Throughput & AVG Traffic & Cell Bandwidth & AVG CQI\\
		\hline
		& 00 - 01 & 17.18 & 86.14 & 5.0 & 11.84\\
		& 01 - 02 & 12.97 & 55.06 & 5.0 & 12.36\\ 
		Cell 999 & 02 - 03 & 19.25 & 50.09 & 5.0 & 12.48\\
		(Dataset-A) & $\vdots$ & $\vdots$ &$\vdots$ &$\vdots$ &$\vdots$\\
		& 23 - 00 & 7.17 & 128.63 & 5.0 & 11.84\\ \cline{2-6}
		& Daily AVG & 17.78 & 105.52 & 5.0 & 11.82 \\\hline
		
		
		& 00 - 01 & 17.45 & 16.31 & 5.0 & 11.46\\
		& 01 - 02 & 24.15 & 11.59 & 5.0 & 10.51\\
		Cell 1720 & 02 - 03 & 32.26 & 7.36 & 5.0 & 9.49\\
		(Dataset-A) & $\vdots$ & $\vdots$ &$\vdots$ &$\vdots$ &$\vdots$\\
		& 23 - 00 & 27.62 & 18.38 & 5.0 & 11.19\\ \cline{2-6}
		& Daily AVG & 24.42 & 18.08 & 5.0 & 10.24 \\\hline
		
		& 00 - 01 & 26.56 & 15.65 & 5.0 & 11.51\\
		& 01 - 02 & 37.82 & 7.52 & 5.0 & 10.97\\ 
		Cell 997 & 02 - 03 & 35.80 & 4.62 & 5.0 & 11.53\\
		(Dataset-A) & $\vdots$ & $\vdots$ &$\vdots$ &$\vdots$ &$\vdots$\\
		& 23 - 00 & 31.41 & 22.94 & 5.0 & 11.67\\ \cline{2-6}
		& Daily AVG & 31.05 & 26.72 & 5.0 & 11.48 \\\hline
		
		& 00 - 01 & 16.83 & 4.35 & 5.0 & 10.10\\
		& 01 - 02 & 22.67 & 7.27 & 5.0 & 9.90\\ 
		Cell 1284 & 02 - 03 & 22.94 & 4.99 & 5.0 & 9.75\\
		(Dataset-B) & \vdots & \vdots &\vdots &\vdots &\vdots\\
		& 23 - 00 & 28.62 & 5.76 & 5.0 & 10.45\\ \cline{2-6}
		& Daily AVG & 17.92 & 7.18 & 5.0 & 10.16 \\\hline
		
		& 00 - 01 & 8.12 & 6.28 & 3.0 & 8.95\\
		& 01 - 02 & 18.22 & 5.73 & 3.0 & 9.12\\ 
		Cell 1127 & 02 - 03 & 18.66 & 6.00 & 3.0 & 9.24\\
		(Dataset-B) & \vdots & \vdots &\vdots &\vdots &\vdots\\
		& 23 - 00 & 14.91 & 7.61 & 3.0 & 8.96\\ \cline{2-6}
		& Daily AVG & 11.09 & 10.38 & 3.0 & 8.98 \\\hline
		
		& 00 - 01 & 32.50 & 29.14 & 5.0 & 10.35\\
		& 01 - 02 & 30.39 & 22.81 & 5.0 & 10.58\\ 
		Cell 986 & 02 - 03 & 33.88 & 22.67 & 5.0 & 10.99\\
		(Dataset-B) & \vdots & \vdots &\vdots &\vdots &\vdots\\
		& 23 - 00 & 29.25 & 37.42 & 5.0 & 10.21\\ \cline{2-6}
		& Daily AVG & 22.45 & 43.41 & 5.0 & 10.32 \\\hline
	\end{tabular}
    }
    \vspace{-.2cm}
	\caption{Samples from Dataset-A and B.}
	\vspace{-.5cm}
	\label{tab:dataset_A}
\end{table}

		
		
		

\subsection{Reinforcement Learning Modeling} \label{sec:appendix_rl}
As we mentioned in Sec.~\ref{sec:simulation}, for problems similar to ours which require following a policy to choose an action to interact with an environment, reinforcement learning (RL) modeling is a good candidate. 
Among all the RL methods, the actor-critic (AC) method is a potential match to our problem due to its continuous action space and modeling flexibility.
AC methods have been widely used in cellular network problems and for different tasks such as power allocation and traffic forecasting \cite{AC_traffic, AC_thrp, AC_iot, AC_resal}.  
Hence, as an additional benchmark we compare our proposed algorithm with a basic AC algorithm we implemented to adapt to our problem where the actor model recommends actions in order to maximize the reward predicted by the critic model.

We use TAG-GCN as the critic model to predict the reward which can be either throughput ratio or throughput to train the actor model. 
For the actor model, we use the same modules presented for TAG-GCN to do the neighbor auto-grouping and capture the temporal dependencies in the data and train it with the following loss function
\begin{equation}
    { {Loss (actor)}}=  \frac{1}{T}\sum_{t=1}^T\frac{1}{N_t}\sum_{v=1}^{N_t} \sqrt{\big|1-\hat{\beta}_t^{v}(\hat{a}_t^v)\big|} + \lambda_3||\Theta_3||^2,
\end{equation}
where $\hat{a}_t^v$ is the recommended action by the actor model and $\hat{\beta}_t^{v}(\hat{a}_t^v)$ is the predicted reward by the critic model using the recommended action.  
$\lambda_3$ is the hyperparameter chosen for regularization and ${\Theta}_3$ represents all the trainable parameters in the actor model.

\begin{figure}[t!]
    \centering
    \includegraphics[width=.475\textwidth]{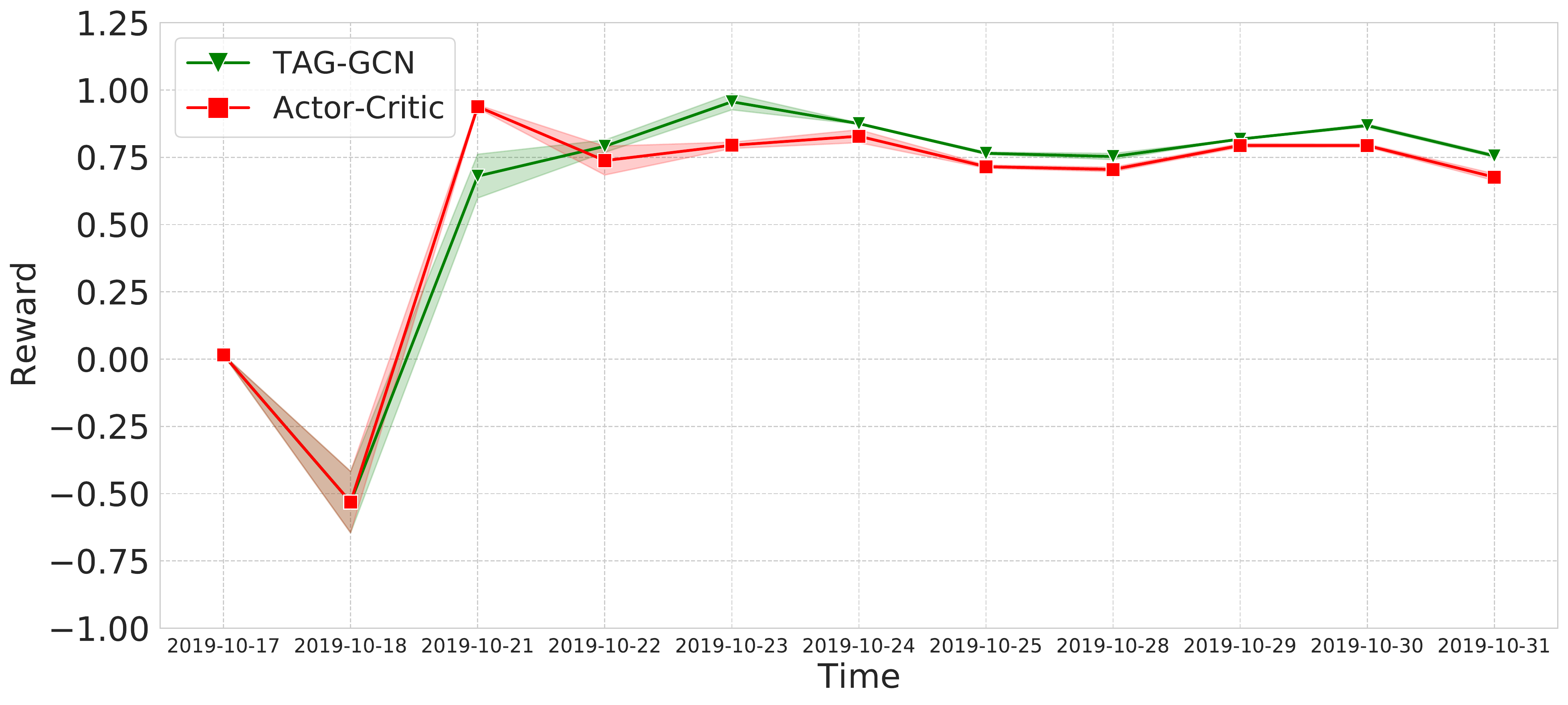}
    \vspace{-.8cm}
    \caption{Performance comparison between our proposed design and actor-critic design for action exploration.}
    \vspace{-.2cm}
    \label{fig:AC}
\end{figure}

Fig.~\ref{fig:AC} shows that TAG-GCN outperforms this basic RL model. The proposed AC model is a very preliminary one and obviously it can be improved even more which is out of the scope of this paper. 
However, this potential improvement comes with the cost of increasing complexity and requires a highly diverse data and more interactions with the environment.
The RL approach generally requires frequent and plenty of interactions with the environment. However, in our case, the action is configured once a day, and the total exploration duration only consists of small number of days, and is far less than what is generally required by the RL models.
Furthermore, RL models usually consider accumulated returns, i.e., impact of the current action on the future performance. However, in our scenario, the current action only affects the loads distribution of the current hour, and does not affect the future performance. Therefore the standard RL approach does not apply here.

\subsection{Alternative reward model comparison} \label{sec:benchmark}
We summarize the alternative designs for the reward model we consider in our paper in Table \ref{tab:Models}.

\begin{table}[t]
    \centering
    {\fontsize{7}{8}\selectfont
    \begin{tabular}{|M{1.5cm}||M{.7cm}|M{1.9cm}|M{2.1cm}|}
     \hline
     \multicolumn{4}{|c|}{Model List} \\
     \hline \centering
     Model Name & Graph & Heterogeneous Graph & Daily Temporal Features\\
     \hline
     MLP  & \xmark & \xmark & \xmark\\
     GCN  & \cmark & \xmark & \xmark\\
     AG-GCN  & \cmark & \cmark & \xmark\\
     TAG-GCN  & \cmark & \cmark & \cmark\\
     \hline
    \end{tabular}}
    \vspace{-.2cm}
    \caption{Properties of Different Models}
    \vspace{-.3cm}
    \label{tab:Models}
\end{table}

\subsection{Additional Model Evaluation Metrics  (avg. Rank)}
We present the average rank of different models across different evaluation days in terms of the achieved test MSE for both throughput ratio and cell throughput models in Dataset-A and B, in Table~\ref{tab:rankA} and \ref{tab:rankB}, respectively.
As seen, considering the results of both datasets, TAG-GCN can achieve better performance than other proposed model which is due to its capacity to learn the spatial and temporal dependencies in the network.

\begin{table}[t!]
    \centering
    {\fontsize{9}{10}\selectfont
    \begin{tabular}{|c|c|c|c|}
        \hline
         Model Name & Ratio Model & Throughput Model & Average\\ \hline
         MLP & 3.83 & 3.83 & 3.83\\
         GCN & 3.17 & 3.00 & 3.08\\
         AG-GCN & 1.67 & 1.67 & 1.67\\
         TAG-GCN & \textbf{1.33} & \textbf{1.50} & \textbf{1.42}\\ \hline
    \end{tabular}}
    \vspace{-.3cm}
    \caption{Average rank of the learning models in terms of the achieved test MSE across the evaluation days for Dataset-A}
    \vspace{-.2cm}
    \label{tab:rankA}
\end{table}

\begin{table}[t!]
    \centering
    {\fontsize{9}{10}\selectfont
    \begin{tabular}{|c|c|c|c|}
        \hline
         Model Name & Ratio Model & Throughput Model & Average\\ \hline
         MLP & 3.50 & 3.83 & 3.67\\
         GCN & 3.00 & 3.00 & 3.00\\
         AG-GCN & 2.00 & \textbf{1.50} & 1.75\\
         TAG-GCN & \textbf{1.50} & 1.67 & \textbf{1.58}\\ \hline
    \end{tabular}}
    \vspace{-.3cm}
    \caption{Average rank of the learning models in terms of the achieved test MSE across the evaluation days for Dataset-B}
    \vspace{-.4cm}
    \label{tab:rankB}
\end{table}

\subsection{Expert Rule-Based Parameter Optimization Strategy} \label{sec:expertrule}
The experts rule-based method can set the actions for each cell $v\in N_t$ for day $t+1$ based on the load balancing as:
\begin{equation}\label{eq:experts}
    a_t^v = \begin{cases}
r_1^v(\beta_{t}^v-1) &\beta_{t}^v\ge 1\\
r_2^v(\beta_{t}^v-1) &\beta_{t}^v<1
\end{cases},
\end{equation}
where $r_1^v, r_2^v < 0$ are the weights set by human experts based on domain knowledge and assumptions about the network dynamics. 
The intuition is to increase the A2-threshold to release traffic when a cell is overloaded and decrease the value in the opposite case.

\subsection{Additional Experimental Results} \label{sec:ablationStudies}                                              
\begin{figure}[t!]
    \vspace{-.4cm}
    \subfloat[]{\includegraphics[width=.23\textwidth]{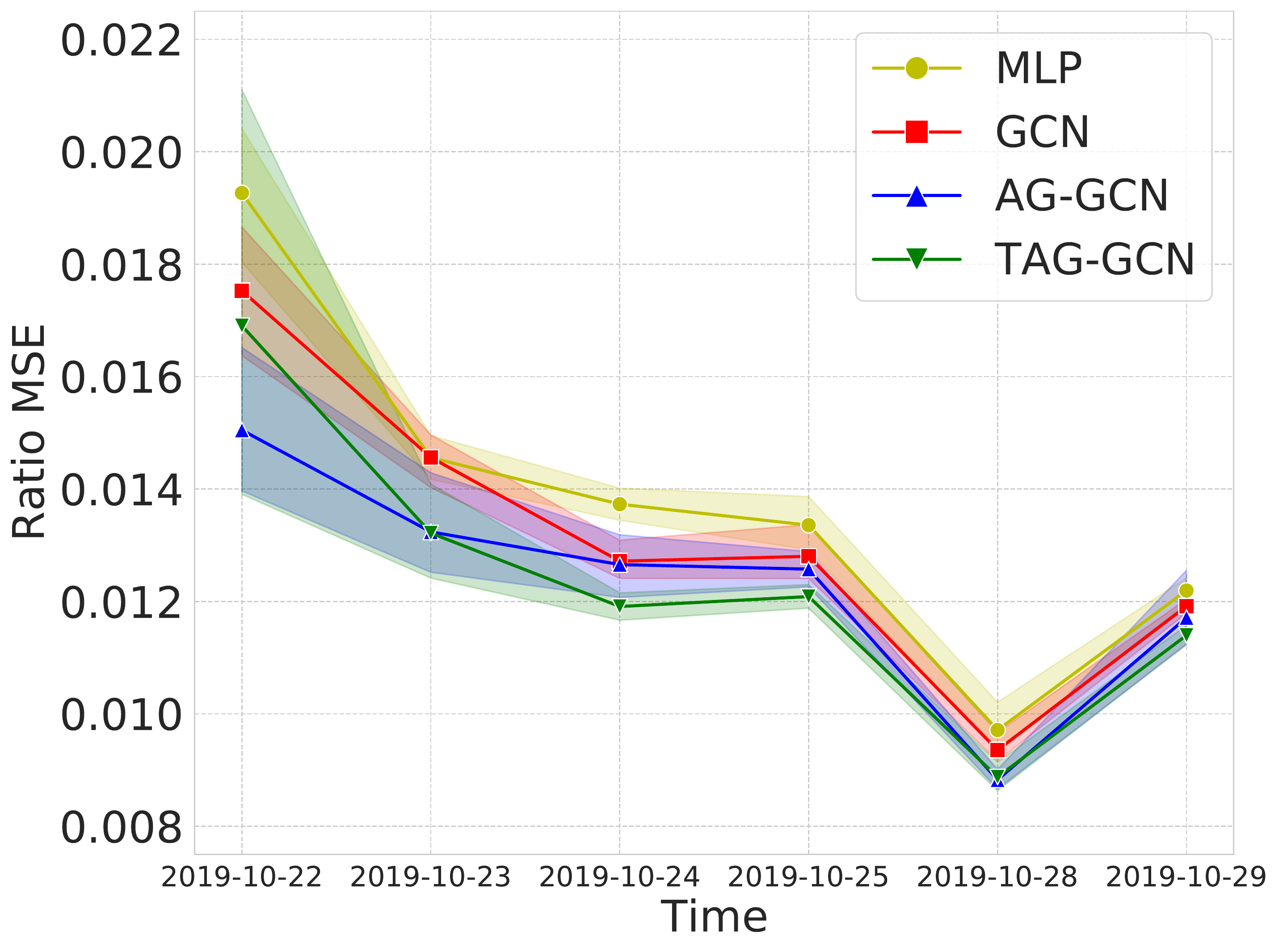}}
	\subfloat[]{\includegraphics[width=.23\textwidth]{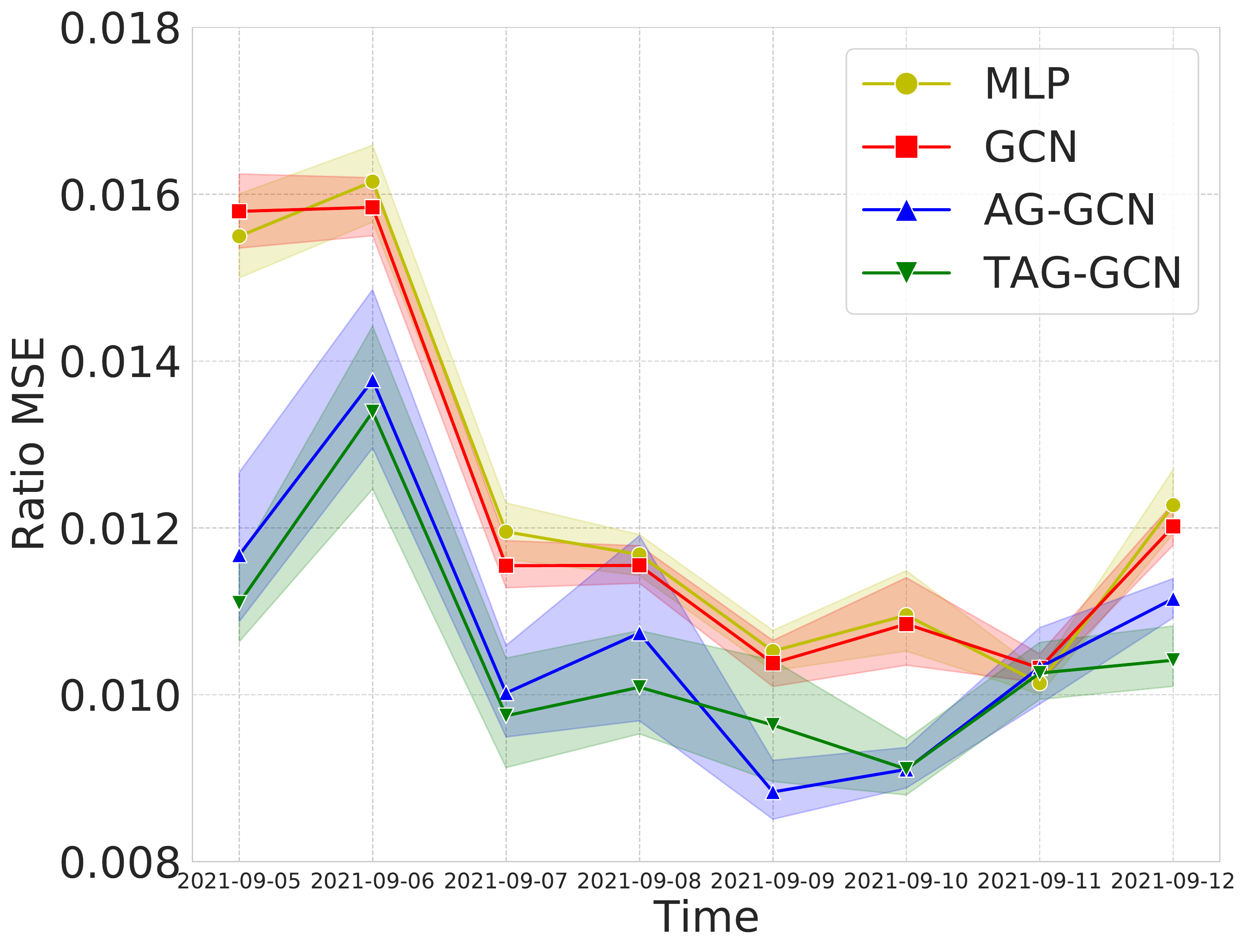}}
	\vspace{-.45cm}
	\caption{Mean squared error of the throughput ratio for test data of (a) Dataset-A and (b) Dataset-B for TAG-GCN.}
	\vspace{-.4cm}
	\label{fig:TAG-GCN_test_mse_ratio}
\end{figure}

\subsubsection{Reward Modeling Performance for Throughput Ratio}\label{sec:appendix_mse}
We repeat the same experiment as Fig.~2 in the paper but for the throughput ratio reward and compare the performance measured in the achieved MSE by different models for Dataset-A and B in Fig.~\ref{fig:TAG-GCN_test_mse_ratio}.

\subsubsection{Ablation Study on the Multi-Objective Action Recommendation Strategy} \label{sec:appendix_multi_obj}

\begin{figure}[t!]
    \centering
    \includegraphics[width=.475\textwidth]{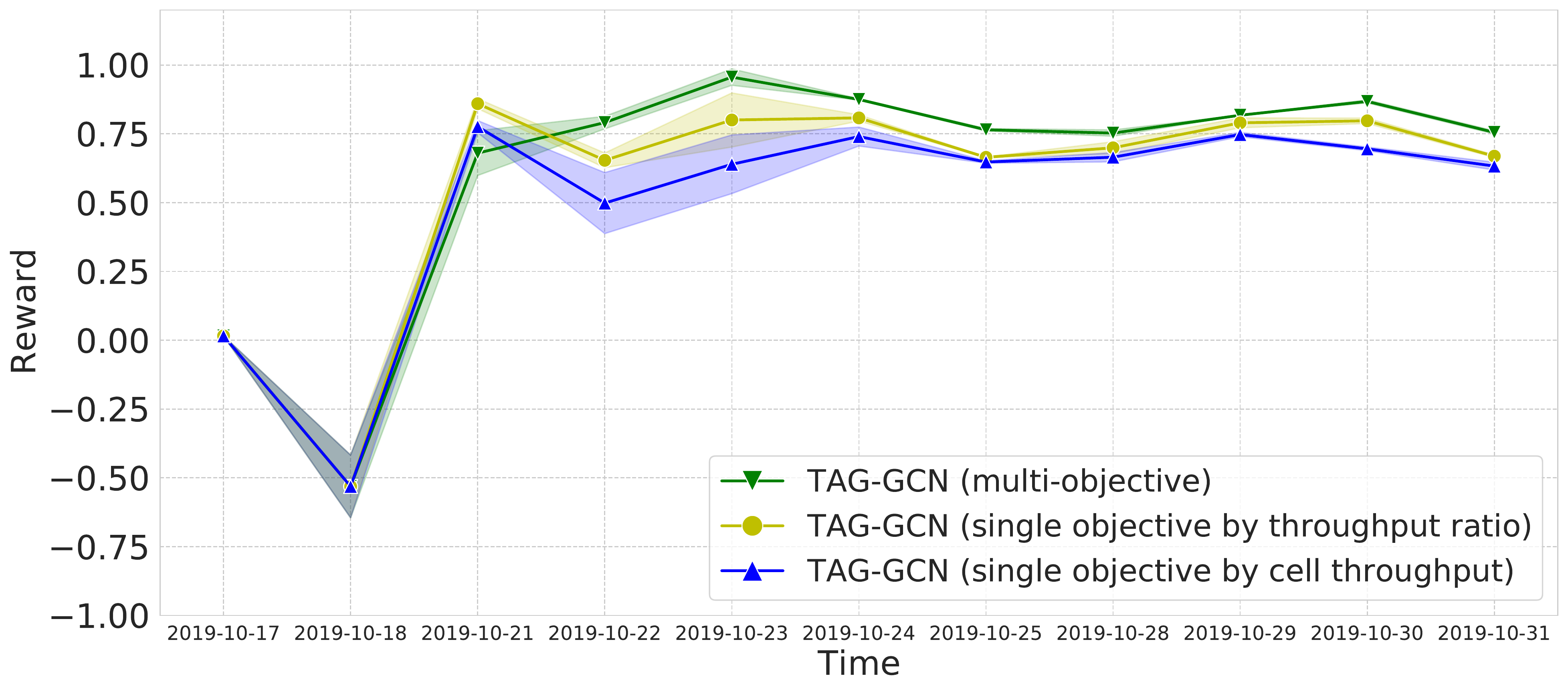}
    \vspace{-.8cm}
    \caption{The impact of different optimization objectives for the parameter configuration in our proposed TAG-GCN model.}
    \vspace{-.3cm}
    \label{fig:Multi_objPerformance}
\end{figure}

In Fig.~\ref{fig:Multi_objPerformance}, we show an ablation study on the effectiveness of our proposed multi-objective action recommendation strategy. 
For each setting, we repeat our experiment five times with TAG-GCN as the reward model but follow different action configuration strategies. Other than the multi-object formulation, we consider two alternatives for action configuration, one by considering only the cell throughput in the optimization problem, and one by considering only the load-balancing objective.
As shown in Fig.~\ref{fig:Multi_objPerformance}, our proposed TAG-GCN model with both load-balancing and throughput as optimization objectives can outperform others, achieving the best average throughput in the final days across random trials.

\subsubsection{The Quantitative Results} \label{sec:appendix_quantitive}

The quantitative results for Fig.~\ref{fig:ModelPerformance} and Fig.~\ref{fig:Multi_objPerformance} are summarized in Table~\ref{tab:num_res_diff_models} and Table~\ref{tab:num_res_diff_opt}, respectively. 
We present the achieved network throughput by different models divided by the performance achieved by the best actions of the simulator for the last 4 days of the experiments. 
It should be noted that in all these experiments the model is trained up to Oct. 29 and after that the model is not updated and only performs the inference step.
As seen in Table~\ref{tab:num_res_diff_models}, the proposed method is outperforming the rest of the methods and yields better average throughput. Furthermore, our method also yields lower variance with respect to change in the initial values of the actions, which is a significant advantage. In general, our method proves to be very robust in this regard.
The second table demonstrates the effectiveness of the multi-objective optimization strategy. 
In this table, we have denoted three different variants of TAG-GCN which represent three optimization strategies, namely, multi-objective strategy using both throughput and throughput ratio models (TAG-GCN1), single objective strategies using throughput ratio (TAG-GCN2), and throughput models (TAG-GCN3). As the results in the table show, the multi-objective strategy outperforms the single objective ones and therefore is superior.

\subsubsection{The Average Rank of Different Models (Fig.~\ref{fig:ModelPerformance})}
As stated earlier the models for the experiment depicted in Fig.~\ref{fig:ModelPerformance} are trained up to Oct. 29 and for the rest of days they are not updated.
In Table~\ref{tab:RewardAVGRank}, we compare the performance of the proposed learning-based models in terms of the achieved average rank.
We separate the results for before and after Oct. 29 to show the generalization of different methods.
As seen TAG-GCN has the best performance both before and after freezing the learning model.



\begin{table}[t!]
    \centering
    \resizebox{\columnwidth}{!}{
    \begin{tabular}{|c|c|c|c|c|} \hline
         Model Name &  \textit{10-28} & \textit{10-29} & \textit{10-30} & \textit{10-31}\\ \hline
         Experts Rule-Based & $.47\pm.0026$ & $.37\pm.0173$ & $.62\pm.0007$ & $.41\pm.0090$\\
         MLP & $.68\pm.0073$ & $.76\pm.0080$ & $.87\pm.0102$ & $.66\pm.0110$\\
         GCN & $.66\pm.0148$ & $.76\pm.0077$ & $.86\pm.0002$ & $.65\pm.0118$\\
         AG-GCN & $.68\pm.0345$ & $.77\pm.0055$ & $.86\pm.0019$ & $.67\pm.0068$\\
         TAG-GCN & $\mathbf{.71\pm.0136}$ & $\mathbf{.78\pm.0005}$ & $\mathbf{.93\pm.0074}$ & $\mathbf{.73\pm.0069}$ \\ \hline
    \end{tabular}}
    \vspace{-.3cm}
    \caption{The Average throughput divided by the throughput achieved from the best action designed by the simulator (in last four days) for different reward models.}
    \vspace{-.2cm}
    \label{tab:num_res_diff_models}
\end{table}

\begin{table}[t!]
    \centering
    \resizebox{\columnwidth}{!}{
    \begin{tabular}{|c|c|c|c|c|} \hline
         Model Name &  \textit{10-28} & \textit{10-29} & \textit{10-30} & \textit{10-31}\\ \hline
         TAG-GCN1 & $\mathbf{.71\pm.0136}$ & $\mathbf{.78\pm.0005}$ & $\mathbf{.93\pm.0074}$ & $\mathbf{.73\pm.0069}$ \\
         TAG-GCN2 & $.66\pm.0259$ & $.75\pm.0212$ & $.85\pm.0131$ & $.65\pm.0060$\\
         TAG-GCN3 & $.63\pm.0199$ & $.71\pm.0094$ & $.74\pm.0089$ & $.61\pm.0170$\\
         \hline
    \end{tabular}}
    \vspace{-.3cm}
    \caption{The 95\% confidence interval for the average throughput divided by the throughput achieved from the best action designed by the simulator (in last four days) for different optimization methods using TAG-GCN model.}
    \label{tab:num_res_diff_opt}
    \vspace{-.2cm}
\end{table}

\begin{table}[t!]
    \centering
    \resizebox{\columnwidth}{!}{
    \begin{tabular}{|c|c|c|c|}
    \hline
         Model Name & Before Oct. 29 & After Oct. 29 & All Days\\ \hline
         MLP & 2.80 & 3.67 & 3.13\\
         GCN & 2.80 & 3.33 & 3.00\\
         AG-GCN & 3.20 & 2.00 & 2.75\\
         TAG-GCN & \textbf{1.20} & \textbf{1.00} & \textbf{1.13}\\ \hline
    \end{tabular}}
    \vspace{-.3cm}
    \caption{Average rank of different models in terms of the achieved reward before and after freezing the learning models}
    \vspace{-.2cm}
    \label{tab:RewardAVGRank}
\end{table}

\subsubsection{Different action initialization strategies}\label{sec:appendix_init}
\begin{figure}[t!]
    \vspace{-.2cm}
    \centering
    \includegraphics[width=.475\textwidth]{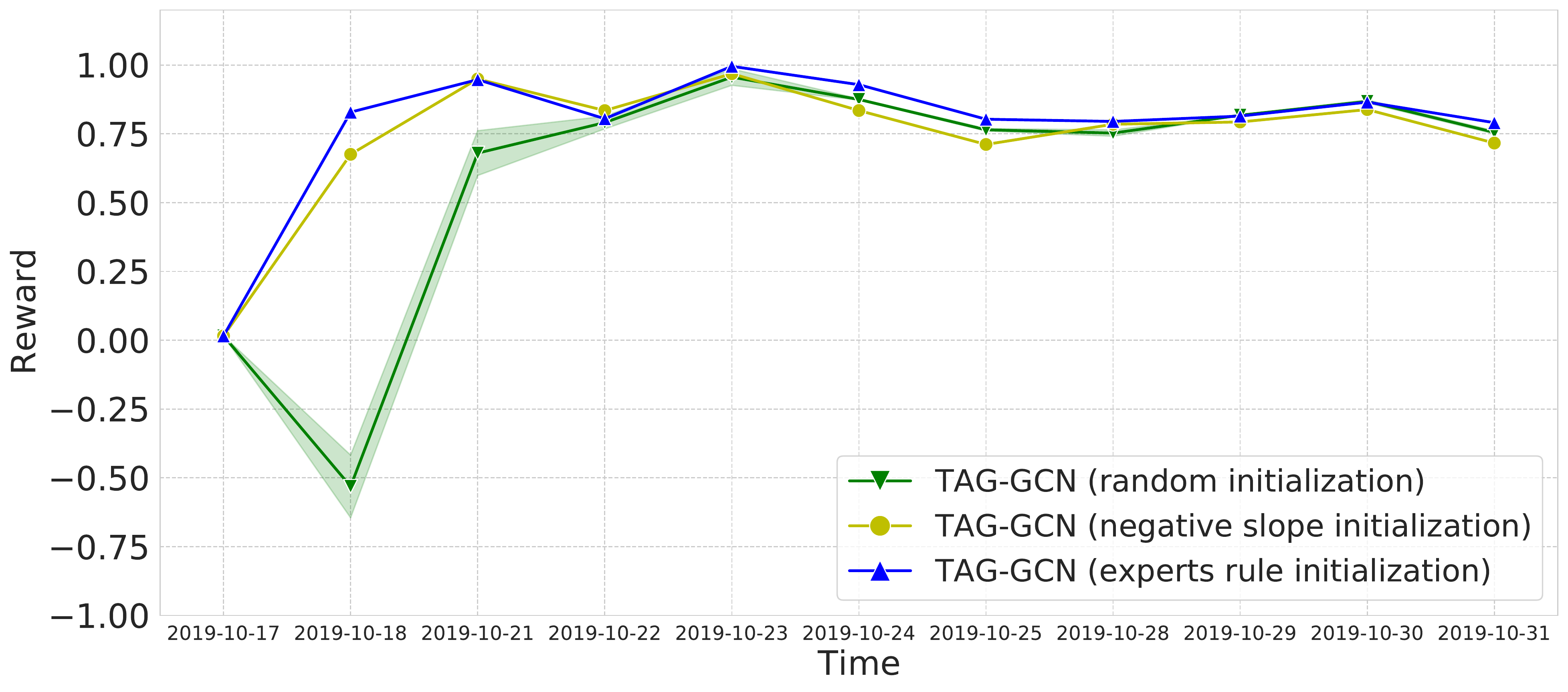}
    \vspace{-.8cm}
    \caption{Performance comparison of the TAG-GCN model under different types of action initialization for Oct. 18.}
    \vspace{-.4cm}
    \label{fig:InitializationPerformance}
\end{figure}

In Fig.~\ref{fig:InitializationPerformance}, we present an ablation study on the different initialization schemes on the first action exploration day.
In addition to the random initialization, we use two more action initialization schemes including the expert rule and the negative slope.
The expert rule initialization is given in (\ref{eq:experts}).
The negative slope procedure for the action initialization is defined based on the throughput ratio of the first day of each cell $v$, i.e., $\beta_{t_1}^{v}$ as 
\begin{equation}
    a_{t_1}^{v} = \Big[\frac{2(\beta_{t_1}^{v} - \rho)\phi}{\rho - \tau} + \phi\Big],
\end{equation}
where $\rho = \min_{v\in N_{t_1}}(\beta_{t_1}^{v})$, and $\tau = \max_{v\in N_{t_1}}(\beta_{t_1}^{v})$ and $[\cdot]$ is the rounding operation.
For controlling the variation of action, the maximum range of change is defined as $\phi$. We can observe that regardless of the type of initialization, the model can converge to almost the same performance in the final days, which indicates that even without strong prior information from the network, our design can still quickly converge to the desired model performance.

\end{document}